# Magnetic Nanoparticles in Nanomedicine

*Kai Wu*[1], *Diqing Su*[2], *Jinming Liu*[1], *Renata Saha*[1], *and Jian-Ping Wang*[1,*]

[1]Department of Electrical and Computer Engineering, University of Minnesota, Minneapolis, Minnesota 55455, USA

[2]Department of Chemical Engineering and Material Science, University of Minnesota, Minneapolis, Minnesota 55455, USA

**E-mail: jpwang@umn.edu*

**Abstract:** Nanomaterials, in addition to their small size, possess unique physicochemical properties that differ from the bulk materials, making them ideal for a host of novel applications. Magnetic nanoparticle (MNP) is one important class of nanomaterials that have been widely studied for their potential applications in nanomedicine. Due to the fact that MNPs can be detected and manipulated by remote magnetic fields, it opens a wide opportunity for them to be used *in vivo*. Nowadays, MNPs have been used for diverse applications including magnetic biosensing (diagnostics), magnetic imaging, magnetic separation, drug and gene delivery, and hyperthermia therapy, etc. This review aims to provide a comprehensive assessment of the state-of-the-art biological and biomedical applications of MNPs. In addition, the development of high magnetic moment MNPs with proper surface functionalization has progressed exponentially over the past decade. Herein, we also reviewed the recent advances in the synthesis and surface coating strategies of MNPs. This review is not only to provide in-depth insights into the different synthesis, biofunctionalization, biosensing, imaging, and therapy methods but also to give an overview of limitations and possibilities of each technology.

Table of Contents

## 1 Introduction

Magnetic nanoparticles (MNPs), with the size between 1 nm and 100 nm, are one important nanomaterial for science and technology in the past two decades. The unique characteristics of MNPs, such as high surface to volume ratio and size-dependent magnetic properties, are drastically different from those of their bulk materials. MNPs have been receiving tremendous attention in multiple areas such as data storage, spintronics, catalyst, neural stimulation, and gyroscopic sensors, etc.[1-20] Nowadays, many methods have been developed to fabricate MNPs due to their wide applications, and there are mainly two approaches to obtain MNPs: top-down approach and bottom-up approach. MNPs with a properly functionalized surface can be physically and chemically stable, biocompatible, and environmentally safe. In addition, biological samples exhibit virtually no magnetic background, thus high sensitivity measurements can be performed on minimally processed samples in MNP-based biological and biomedical applications. Thus, MNPs have been successfully applied as contrast enhancers in magnetic resonance imaging (MRI) [21-26], tracer materials in magnetic particle imaging (MPI) [27-33], tiny heaters in magnetic hyperthermia [34-37], carriers for drug/gene delivery [38-43], tags for magnetic separation [12, 44-49], etc. The ease of synthesis and facile surface chemistry, with comparable size to biologically important ligands, has generated much eagerness in applying MNPs to clinical diagnostics and therapy [50-58].

Herein, we have reviewed different strategies for the synthesis and biofunctionalization of MNPs, as well as the diagnostic and therapeutic applications of MNPs. Our aim is to discuss the challenges of working with MNPs while giving an overall overview of the state-of-the-art.

## 2 Physical Properties of Magnetic Nanoparticles (MNPs)

### 2.1 From Bulk Material to Nanoparticle

MNPs, with comparable sizes to biologically important objects [59], have demonstrated unique properties such as larger surface-to-volume ratio, excellent reactivity, exceptional magnetic response compared to their bulk materials [60, 61]. Materials such as pure metals (Fe, Co, Ni, etc.), alloys (FeCo, alnico, permalloy, etc.), and oxides ($Fe_3O_4$, $\gamma$-$Fe_2O_3$, $CoFe_2O_4$, etc.) with high saturation magnetizations are preferred for the production of MNPs. Although pure metals are able to yield higher saturation magnetizations, they are not suitable for clinical use due to the high toxicity and oxidative properties. One of the most widely used MNP types is iron oxide considering the high chemical and colloidal stability, high biocompatibility, and low cost.

Magnetic properties of MNPs, such as magnetic moment and anisotropy constant, depend strongly on their size and shape. Magnetic moment $(\mu)$ is the product of magnetization $(M)$ and magnetic core volume $V_m$ which is the most important property of MNP for nanomedicine related applications. A higher magnetic moment yields a more pronounced detection signal (for magnetic biosensing and imaging-related applications) as well as a larger magnetic force (for magnetic manipulation and drug/gene delivery related applications) [62, 63]. Due to the lack of translational crystal symmetry in the surface layer (also known as surface spin-canting effect) [64-67], the magnetic properties of surface layer are very different from the interior region of MNP, as a result, decreased saturation magnetizations $M_s$ and increased anisotropy constants are observed in MNPs compared to their corresponding bulk materials. This spin-canting effect is not affected by the organic capping but the magnetic core size of MNP [68] and the particle synthesis methods [69]. It can be understood in terms of a core-shell structure (as shown in Figure 1(a)) where the spins in the magnetic core are fully aligned along the applied magnetic field while the spins in the shell are canted relative to the field. For a spherical MNP, its saturation magnetization can be modeled by [70]:

$$M_s = M_{sb}(1 - 2\delta/D)^3 \quad (1)$$

Where $M_{sb}$ is the saturation magnetization of the bulk material, $D$ is the diameter of the magnetic core, and $\delta$ is the thickness of the spin canting layer. Dutta *et al.* [71] have reported the fitted values of $\delta = 0.68\, nm$ and $M_s = 92\ emu \cdot g^{-1}$ ($\sim 4.76 \times 10^5\ J/m^3T$) for magnetite nanoparticles based on experimental results.

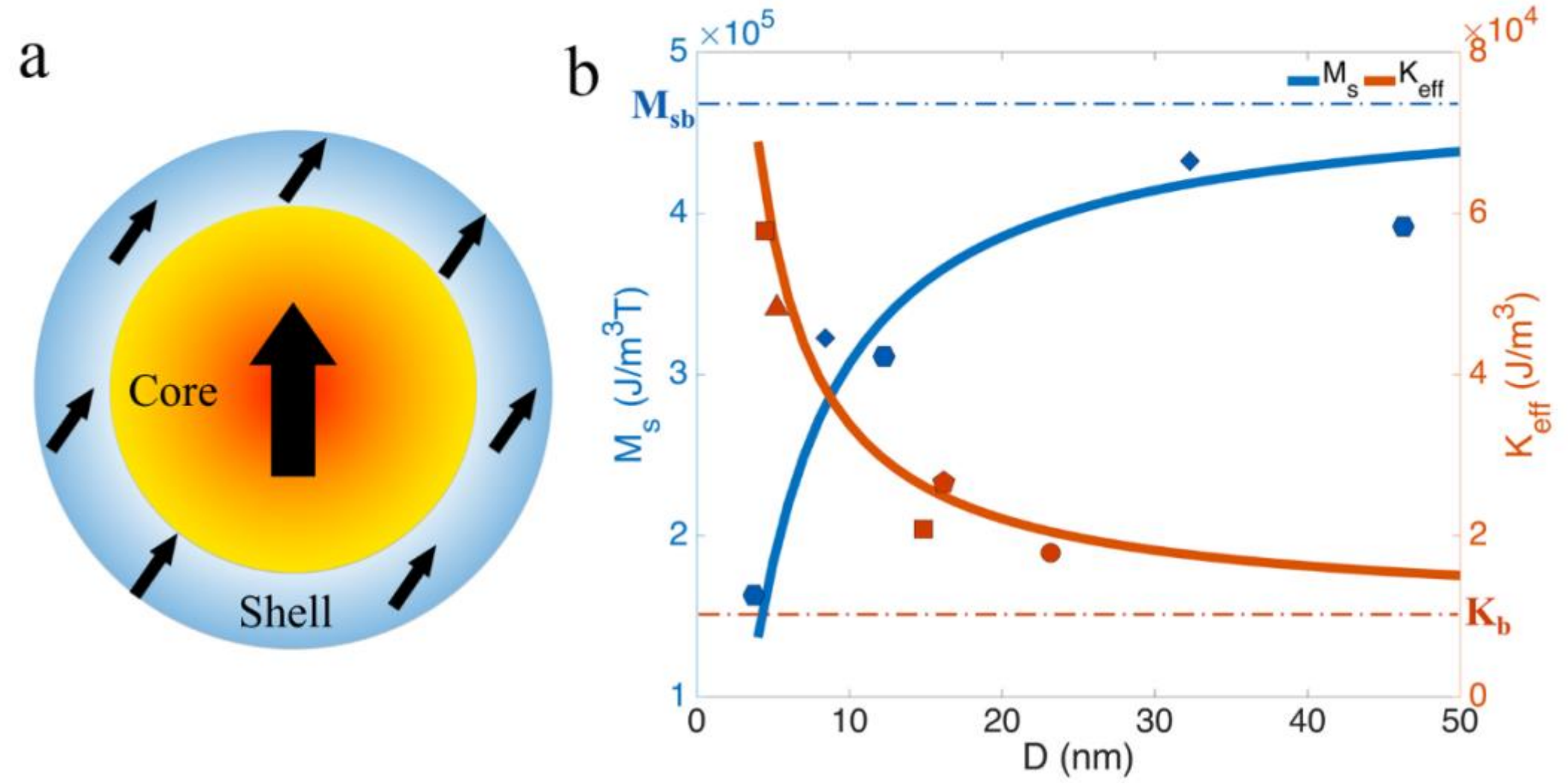

Figure 1. (a) Schematic view of general spin canting geometry, the core-shell model. (b) Theoretical $M_s$ (blue solid line) and $K_{eff}$ (orange solid line) versus magnetite nanoparticle diameter $D$ at 300 K. Horizontal dash-dot lines exhibit $M_{sb}$ and $K_b$, respectively. Theoretical data are compared with experimental ones. Blue diamonds from [72], blue hexagons from [73], orange triangle from [74], orange squares from [75], orange pentagon from [31], and orange circle from [76]. ((b) reprinted with permission from reference [67], copyright 2017 IOP Publishing)

On the other hand, due to the spin-canting effect, the observed anisotropy constants from MNPs are always larger than their corresponding bulk materials [67]. The effective anisotropy constant $K_{eff}$ is modeled by [77]:

$$K_{eff} = K_b + \left(\frac{6\Phi}{D}\right) K_s \ (2)$$

Where, $K_b$ and $K_s$ are the bulk and surface anisotropy constants, respectively. For spherical particles, $\Phi = 1$. Demortiere *et al.* [77] have reported the fitted values of $K_b = 1.04 \times 10^5\ erg \cdot cm^{-3}$ ($\sim 1.04 \times 10^4\ J/m^3$) for bulk magnetite, and $K_s = 3.9 \times 10^{-2}\ erg \cdot cm^{-2}$ ($\sim 3.9 \times 10^{-5}\ \mathrm{J/m^2}$). The effective anisotropy is also affected by the MNP shape (i.e. shape anisotropy), for spherical nanoparticles, shape anisotropy is negligible compared to surface anisotropy.

These two models in equations (1) & (2) have been applied to predict the magnetic properties of magnetite nanoparticles with respect to magnetic core diameters ($D$), and they were examined by comparing with experimental data from literature available (see Figure 1(b)).

## 2.2 From Multi-domain to Single-domain MNPs

Bulk magnetic materials are composed of microscopic crystalline grains (also known as polycrystalline). Each grain is single crystal, with crystal lattices oriented in different directions, has an easy axis of magnetization. To reduce the magnetostatic energy, each grain spontaneously divides into many magnetic domains separated by domain walls, called multi-domain state. The magnetizations of different domains point in different directions in a more or less random manner while the magnetization within each domain points uniformly to one direction. As a single crystal grain splits into multi-domains, the magnetostatic energy reduces but meanwhile extra energy (domain wall energy) is required. Magnetic domains stabilize to a critical size $D_{crit}$ when the energy cost of generating an additional domain wall is equal to the magnetostatic energy saved. This critical size varies for different magnetic materials (proportional to$(A/2K)^{1/2}$) which is determined by the material properties such as exchange stiffness ($A$) and anisotropy constant ($K$), usually, the critical size is between 10 nm and 100 nm. The critical size range is bounded by a lower limit (superparamagnetism) and a higher limit (multi-domain), see Figure 2(a). The critical sizes for the observation of superparamagnetism, single-domain, and multi-domain behaviors of a variety of common ferromagnetic fine nanoparticles can be found from reference [78].

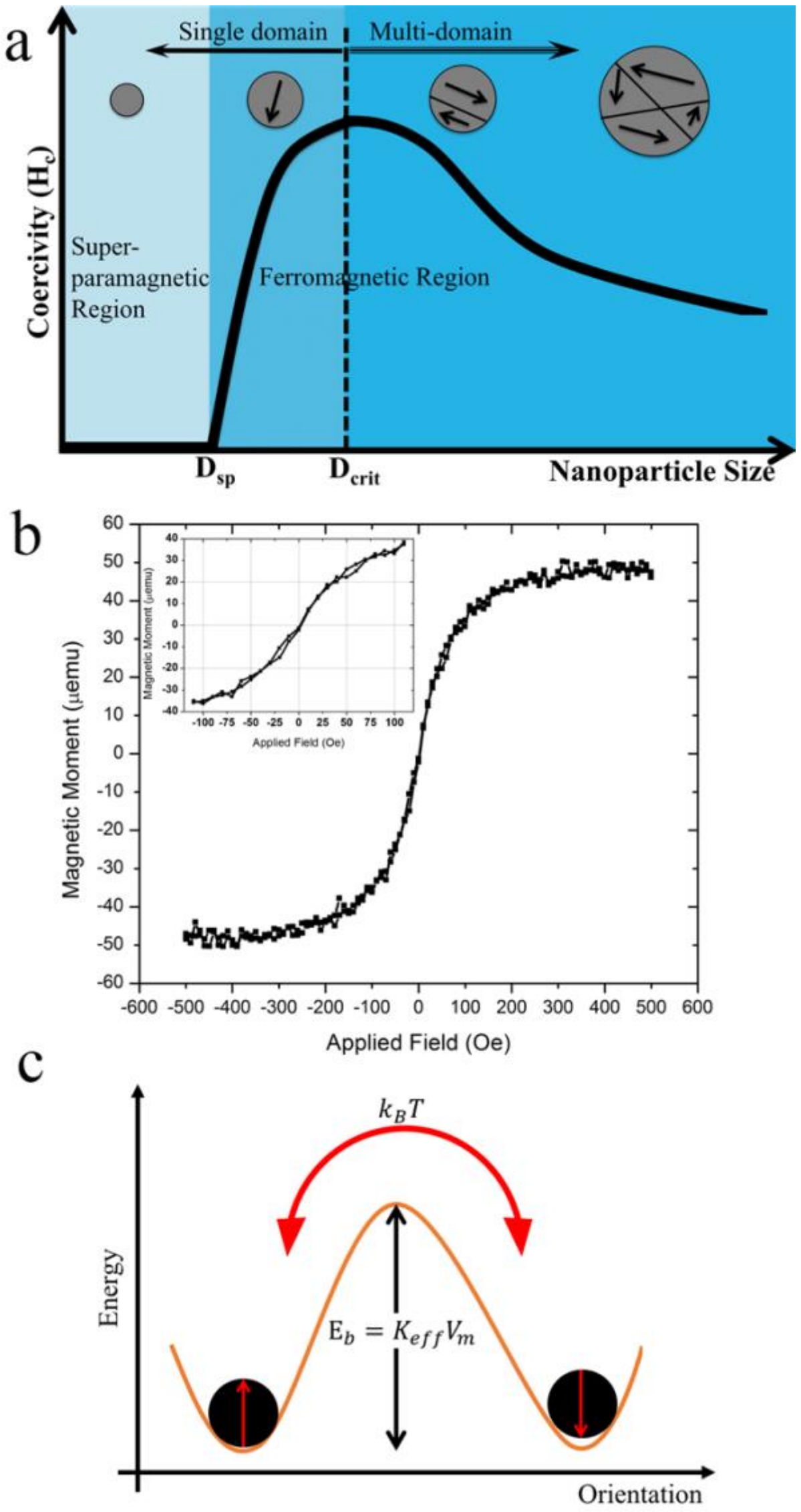

Figure 2. (a) Transition from superparamagnetic to the multi-domain region. The inset figure shows qualitative behaviors of the size-dependent coercivities of MNPs. $D_{sp}$ and $D_{crit}$ are, respectively, the transition sizes from superparamagnetic to a blocked state, and from single-domain to multi-domain regions. (b) Magnetization curve of superparamagnetic nanoparticles measured at room temperature by VSM. Superparamagnetic nanoparticle shows zero magnetic coercivity. (c) Schematic view of energy barrier and thermal fluctuation. ((b) reprinted with permission from reference [79], copyright 2015 AIP Publishing LLC)

Single-domain MNPs are particles with internal magnetizations pointing uniformly in one direction, thus, these particles bear magnetizations equal to their spontaneous magnetization $M_s$, and they have the largest possible magnetic moment of $\mu = V_m M_s$, where $V_m$ is the magnetic core volume of the particle. The magnetization of a single-domain particle rotates as one single giant magnetic moment (called the macro-spin approximation) under external magnetic fields and its hysteresis has been well described by the Stoner-Wohlfarth model.

Due to the magnetic anisotropy, the magnetic moment of a single-domain MNP has two preferred orientations which are antiparallel to each other and are both aligned along its “easy axis”. These two preferred directions are energy minimums separated by an energy barrier of $E_b = K_{eff}V_m$, which prevents the magnetic moment from flipping from one direction to the other. However, thermal fluctuations cause the magnetic moment to rotate in a random manner. As is shown in Figure 2(c), under a finite temperature $T$, if the energy barrier $E_b$ is comparable to or smaller than the thermal fluctuation energy $k_B T$, the magnetic moment jumps from one direction to the other frequently during a measurement time period of $\tau_m$, then the observed net magnetization is zero, resulting in superparamagnetism. Under a specific measurement time $\tau_m$ and temperature $T$, there is a critical size $D_{sp}$ at which the transition from single-domain to superparamagnetic nanoparticle occurs. Again, this critical size $D_{sp}$ varies for different magnetic materials, usually in the range of several to several tens nanometers. Due to the fast flipping of magnetic moments in superparamagnetic nanoparticles, they show zero magnetic moments in the absence of an external magnetic field. When an external field is applied, their magnet moments tend to align along the field resulting in nonzero net magnetization, the magnetic moment of superparamagnetic nanoparticle *vs* applied field is a reversible S-shaped curve as shown in Figure 2(b).

## 2.3 Superparamagnetism

At a finite temperature $T$, there is a finite probability that the magnetic moment of a superparamagnetic nanoparticle flips between two preferred directions. The mean time between two flips is called zero-field Néel relaxation time expressed as:

$$\tau_N = \tau_0 exp\left(\frac{K_{eff}V_m}{k_B T}\right) \quad (3)$$

Where, $\tau_0$ is attempt time (attempt period), which is around $10^{-10} \sim 10^{-9}$ seconds depending on the material, and $k_B$ is Boltzmann constant.

It can be seen from equation (3) that the Néel relaxation time is an exponential function of the particle size, and it can vary from nanoseconds for nanoparticles to years for larger particles or bulk materials. Néel relaxation process is the rotation of magnetic moment inside a stationary MNP (see Figure 3(b)).

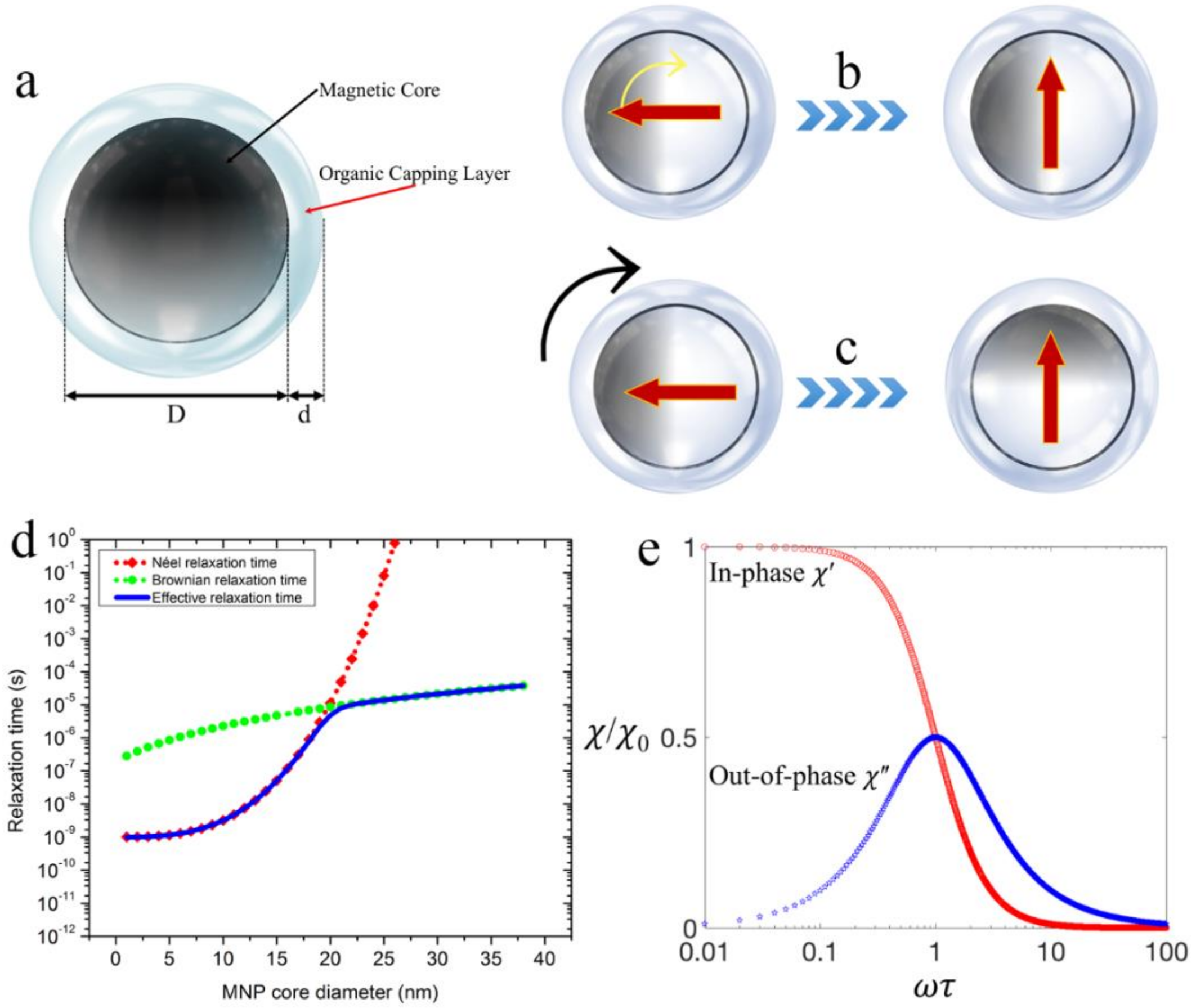

Figure 3. (a) Magnetic core diameter $D$ and organic capping layer with thickness $d$. (b) Néel relaxation process is the rotation of magnetic moment inside a stationary particle. (c) Brownian relaxation process is the rotation of entire particle with its magnetic moment. (d) Néel, Brownian, and effective relaxation time as a function of MNP core diameter, $T = 293\ K$. Viscosity of the MNP solution is assumed to be $\eta = 1\ cp$. (e) The AC magnetic susceptibility has two components: in-phase and out-of-phase. The figure shows the normalized $\chi'$ and $\chi''$ as functions of $\omega\tau$, the out-of-phase component $\chi''$ reaches a maximum when $\omega\tau = 1$. ((d) reprinted with permission from reference [79], copyright 2015 AIP Publishing LLC)

However, for most biomedical applications, superparamagnetic nanoparticles are dispersed in suspensions (also called ferrofluid) where the magnetic relaxivity is a joint effect of both Néel and Brownian processes (see Figure 3(b) & (c)) [67], the zero-field Brownian relaxation time is given as:

$$\tau_B = \frac{3\eta V_h}{k_B T} \quad (4)$$

where $\eta$ is the fluid viscosity, $V_h$ is the hydrodynamic volume of MNP.

The Néel and Brownian relaxation models from equations (3) & (4) are quite simplified, representing a general guideline of non-interacting superparamagnetic nanoparticles under zero magnetic field. Within the assumption of independence of Néel and Brownian processes, the effective relaxation time is given as:

$$\tau = \frac{\tau_N \tau_B}{\tau_N + \tau_B} \quad (5)$$

Figure 2(b) shows the magnetization curve of an ensemble of 25 nm iron oxide superparamagnetic nanoparticles measured by Vibrating Sample Magnetometer (VSM) at room temperature [79]. Since above the blocking temperature $T_B$, the measurement time $\tau_m$ is larger than $\tau_N$, so the nanoparticle appears to be in the superparamagnetic state and its magnetic moment flips several times during one measurement period, thus, the measured magnetization is averaged to zero when the external field $H = 0\ Oe$. As the external magnetic field is applied, the magnetic moments of nanoparticles tend to align with the field direction, resulting in a net magnetization. Hence, superparamagnetic nanoparticles show paramagnetic behavior under small magnetic fields, and the magnetization curve is a reversible S-shape, which is usually simplified by the Langevin model:

$$M(H) = M_S L\left(\mu_0 \frac{\mu H}{k_B T}\right) \quad (6)$$

where $\mu_0$ is the magnetic permeability of vacuum, $L(x)$ is the Langevin function, $H$ is the applied magnetic field.

## 2.4 Magnetic Susceptibility

Magnetic susceptibility $\chi$ is a dimensionless proportionality constant that describes the degree of magnetization of a material in response to external magnetic field, it is the ratio of magnetization $M$ to the field $H$. When subjected to alternating current (AC) magnetic field, the magnetization of MNP may not be able to follow the AC field due to its finite rate of magnetic relaxation, thus, a phase delay between the AC field and the magnetization is introduced. This property introduces a complex magnetic susceptibility $\chi(\omega)$, which can be calculated by the Debye model [80]:

$$\chi(\omega) = \frac{\chi_0}{1 + j\omega\tau} = \chi' + j\chi'' = |\chi| e^{j\varphi} \quad (7)$$

and

$$\chi' = \frac{\chi_0}{1 + (\omega\tau)^2} \quad (8)$$

$$\chi'' = \frac{\chi_0 \omega\tau}{1 + (\omega\tau)^2} \quad (9)$$

$$\varphi = tan^{-1}(\omega\tau) \quad (10)$$

where $\chi_0$ is the DC (direct current) field susceptibility, $\omega$ is the angular frequency of AC field, $\chi'$ and $\chi''$ are the in-phase and out-of-phase components, $\varphi$ is the phase delay.

Figure 3(e) shows the normalized $\chi'$ and $\chi''$ as functions of $\omega\tau$, the out-of-phase component $\chi''$ reaches a maximum when $\omega\tau = 1$, this property is exploited for monitoring the binding of target biomolecules to MNPs in magnetic relaxometry based biosensors. Furthermore, $\chi''$ also holds great significance in magnetic hyperthermia applications. The power generation $P$ (also called specific absorption rate, SAR) by MNPs under an AC magnetic field is defined by Rosensweig theory (RT) [81]:

$$P = \frac{1}{2}\mu_0 \omega \chi'' {H_0}^2 \quad (11)$$

where $H_0$ is the magnitude of AC magnetic field $\boldsymbol{H}$. This equation leads to a global maximum of $P$ when $\omega\tau = 1$, which defines the critical frequency for the system [82]. SAR is a parameter commonly used to characterize the goodness of a given combination of colloidal suspension and magnetic field characteristics to convert the magnetic field energy into thermal energy.

## 2.5 Magnetic Relaxivity

The element that contains an odd number of protons and/or neutrons in its nucleus, such as $^{1}H$, $^{2}H$, $^{13}C$, $^{14}N$, $^{15}N$, $^{17}O$, $^{23}Na$, $^{31}P$, etc., exhibits intrinsic magnetic moment (namely, spin), which is the primary origin of the magnetic resonance signal. Single proton hydrogen $^{1}H$ is one particularly favorable element for nuclear magnetic resonance (NMR) and magnetic resonance imaging (MRI) applications due to its high intrinsic sensitivity and high abundance in water and lipid molecules. For example, the magnetic resonance signals from water or fat within the patient's tissue are monitored for MRI applications, these magnetic resonance signals come from the $^{1}H$ which is abundant in water and lipid molecules [83]. Although NMR has relatively low sensitivity, the MNP provides inherent signal amplification to NMR since each MNP cluster affects billions of adjacent water protons. Thus, MNPs have been widely used as contrast agents in NMR and MRI applications [84, 85].

Under an externally applied static magnetic field ($\boldsymbol{H}_0$ along the $\boldsymbol{z}$-direction), the water proton nuclei align parallel to $\boldsymbol{H}_0$ and precess with the Larmor frequency. As shown in Figure 4(a), when a resonant radio frequency (RF) pulse is applied perpendicular to $\boldsymbol{H}_0$, these nuclei are excited to antiparallel states. Upon the removal of RF pulse, these nuclei relax to equilibrium states. In the presence of MNPs, the dipolar magnetic field increases the local field inhomogeneity. When water molecules diffuse to the periphery of MNPs, the coherent precessions of water proton spins are perturbed. As a result, the net effect is a change of magnetic resonance signal, which can be measured as the shortening of the longitudinal and transverse relaxation time of surrounding water proton spins. The longitudinal relaxation time $T_1$ (also known as spin-lattice, or relaxation in the z-direction), as shown in Figure 4(b), is a measure of the time taken for the $z$ component of the nuclear spin magnetization, $M_z$, to return to its thermal equilibrium value $M_0$:

$$M_z(t) = M_0 - [M_0 - M_z(t=0)] \cdot e^{-t/T_1} \quad (12)$$

where the $z$ component magnetization recovers to 63% of its equilibrium value within a time period of $T_1$.

On the other hand, the transverse relaxation time $T_2$ (also known as spin-spin, or relaxation in the x-y plane), as shown in Figure 4(c), is the decay constant of the net magnetization $M_{xy}$ (magnetization perpendicular to $H_0$):

$$M_{xy}(t) = M_{xy}(t=0) \cdot e^{-t/T_2} \quad (13)$$

where the transverse magnetization drops to 37% of its original magnitude within a time period of $T_2$.

Correspondingly, the longitudinal and transverse relaxation rates are reciprocals of $T_1$ and $T_2$:

$$R_1 = \frac{1}{T_1} \quad (14)$$

$$R_2 = \frac{1}{T_2} \quad (15)$$

Magnetic relaxivity is the intrinsic property of MNP contrast agent, it reflects the MNP's ability to increase the longitudinal and transverse relaxation rates of the surrounding nuclear spins, denoted as $r_1$ and $r_2$, respectively:

$$r_1 = \frac{\Delta R_1}{\Delta C} \quad (16)$$

$$r_2 = \frac{\Delta R_2}{\Delta C} \quad (17)$$

Where, $C$ is the concentration of MNPs. Thus, the relaxivities $r_1$ and $r_2$ have units of $nM^{-1} \cdot s^{-1}$.

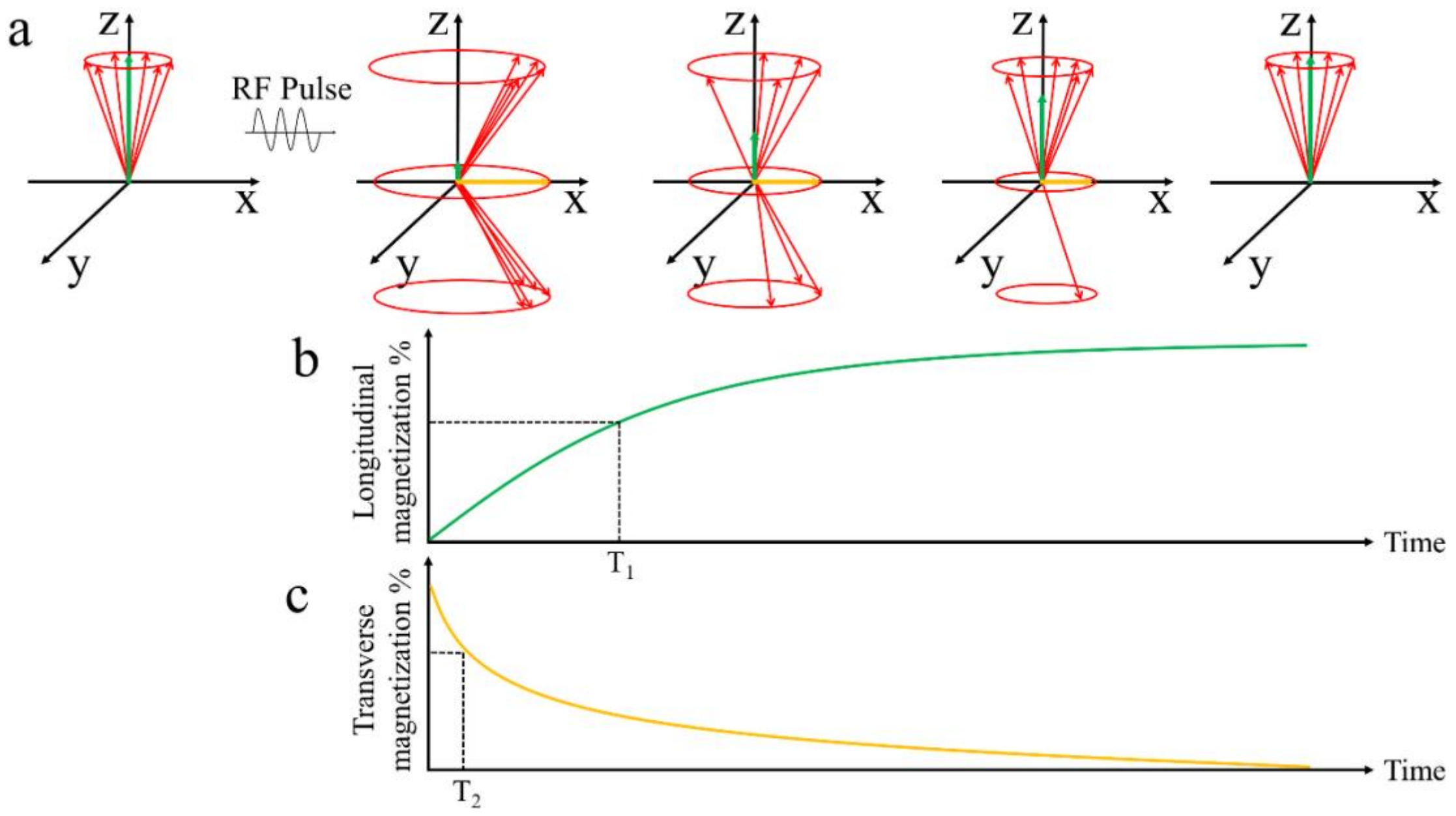

Figure 4. (a) The transition from parallel to antiparallel states upon the application of RF pulse. (b) Longitudinal relaxation time $T_1$. (c) Transverse relaxation time $T_2$.

The sensitives of NMR and MRI are largely dependent on the relaxivities of the MNP contrast agents. Generally, the transverse relaxivity values $r_2$ of MNPs are greater than longitudinal relaxivity $r_1$, thus MNPs are mainly used as a $T_2$-modulating agent and MNPs with higher $r_2$ are preferred for these applications [86]. In addition, the relaxivity is also dependent on the magnitude of $H_0$ [87], temperature, and solvent (e.g., blood, water, plasma). Magnetic properties of MNPs such as saturation magnetization, size [88], and shape [89-91] are reported to affect the relaxivity. Furthermore, the aggregated and evenly dispersed MNPs can also lead to a $T_2$ variation. In real NMR experiments, the measured transverse relaxation time (denoted as $T_2^*$) is always less than or equal to $T_2$, which arises from the inhomogeneities in the main magnetic field.

In recent years, lots of work have been carried out to increase the $r_2$ of contrast agents in order to improve the sensitivity of NMR and MRI [92-94]. In a recent report by Zhou *et al.* [90], they artificially introduced local magnetic field inhomogeneity by clustering of MNPs as well as designing MNPs with heterogeneous geometries (e.g., size and shape) to enhance the $r_2$. Mohapatra *et al.* [95] have reported iron oxide nanorods with very high $r_2$ relaxivity value of $608\, mM^{-1}s^{-1}$, which is mainly attributed to the higher surface area and anisotropic morphology. Furthermore, alloy-based nanomaterials are good candidates for developing $T_2$ contrast agents with high $r_2$ relaxivity values, graphene oxide-$Fe_3O_4$ (GO-$Fe_3O_4$) MNP composite [94] and $Zn^{2+}$ doping controlled MNPs [96] have been reported to effectively increase the $r_2$ value. The classical outer-sphere relaxation theory points out that $r_2/r_1$ increases with increasing particle sizes [97], thus, larger MNPs or MNP clusters are more likely to be better $T_2$-modulating agents.

## 2.6 Dipolar Interactions

MNPs have been successfully applied in magnetic biosensing (i.e. diagnosing), trapping, and therapeutic platforms. Most of these applications are based on the magnetic properties of MNPs, which may vary depending on the MNP aggregation state, namely, the interparticle distances. In a cluster of MNPs, the dipolar interactions (also known as dipole-dipole interactions or dipolar coupling) affect substantially to the collective magnetic behaviors [67, 82, 98]. To be specific, this dipolar interaction modifies the magnetic relaxation time [67, 82], susceptibility, remanence, coercivity, and blocking temperature [99] of MNPs. As a result, the performance of hyperthermia therapy [82, 100, 101] and magnetic particle imaging (MPI) [102] can be drastically modified. As the MNP concentration increases, the interparticle distance decreases, and the dipolar interaction increases, which alters the magnetic response of the whole ferrofluid. For MNPs under an applied magnetic field $\boldsymbol{H}$, the total magnetic field acting on one MNP is the sum of external magnetic field $\boldsymbol{H}$ and the dipolar field $\boldsymbol{H}_{dip}$ which is generated by surrounding MNPs:

$$\boldsymbol{H}_{dip} = \frac{1}{4\pi}\sum_{i\neq j}\frac{3(\boldsymbol{\mu}_j\cdot\boldsymbol{e}_{ij})\cdot\boldsymbol{e}_{ij}-\boldsymbol{\mu}_j}{{r_{ij}}^3} \quad (18),$$

Where, $\boldsymbol{e}_{ij}$ is the unitary vector joining two MNPs, $r_{ij}$ is the distance between these two MNPs. This dipolar field is inversely proportional to the cube of interparticle distance.

As is mentioned in section 2.1, the organic capping layer will not affect the magnetic properties of MNPs, which suggests a possibility of tailoring the interparticle distances by controlling the thickness of capping layer. This organic capping layer can effectively prevent direct surface contact and increase the average interparticle distance thus, reducing the dipolar interactions.

## 2.7 Multicore MNPs

Superparamagnetic NPs show zero remanent magnetization due to thermal fluctuation, as the NP size increases they become ferrimagnetic and show hysteresis. For many MNP-based medical and biological applications, larger

MNPs are preferred to yield higher specific heating losses and higher magnetic moments. However, the non-zero remanent magnetization of larger MNPs leads to agglomeration in the absence of external magnetic field even with polymer coatings. Once the MNP agglomerates reach to the size of a red blood cell (which is around 6 μm), they are at risk of blocking blood vessels and cause dangerous side effects to the patients [103]. To prevent agglomeration, the multicore MNPs (also called superferrimagnetic multicore NPs, i.e., MCNPs) are proposed, they are clusters of smaller superparamagnetic NPs embedded in a polymer matrix. Typically, these multicore MNPs are between 20 and 100 nm and consist of superparamagnetic NPs about the size of around 10 nm. Multicore MNPs show much smaller remanent magnetization compared to the single core MNPs with similar sizes, which could effectively prevent the agglomeration. Multicore MNPs have shown excellent potential for applications in magnetic diagnostics and therapy. Lartigue *et al.* [104] reported magnetically cooperative multicore MNPs with enhanced hyperthermia efficiency and MRI $T_1$ and $T_2$ contrast effects. By applying electrostatic colloidal sorting while preserving a superparamagnetic-like behavior, they have successfully enhanced magnetic susceptibility and decreased surface disorder of multicore MNPs. Kratz *et al.* [105] presented a novel aqueous synthesis for generating multicore MNPs that are suited for both MRI and MPI and allows the combination of these two techniques for biomedical imaging. Lai *et al.* [106] presented multicore $MnFe_2O_4$@$SiO_2$@Ag MNPs with both magnetic and plasmonic properties, which holds great promise for applications in magnetic/photo-thermal hyperthermia and surface-enhanced Raman spectroscopy.

## 3 Synthesis of MNPs

Overall, there are two main approaches to prepare nanoparticles: top-down approach and bottom-up approach. The schematic drawing of these two approaches is shown in Figure 5. In the top-down approach, such as lithography and ball milling, bulk materials (or thin films) are broken down into micrometer or nanometer size. In the bottom-up approach, however, nanoparticles form from atoms followed by nucleation and growth process. There are several bottom-up methods used to prepare nanoparticles, such as gas-phase condensation, chemical vapor deposition, and wet chemical method. In this section, we will focus on these two methods: ball milling method as top-down approach and gas-phase condensation method as bottom-up approach.

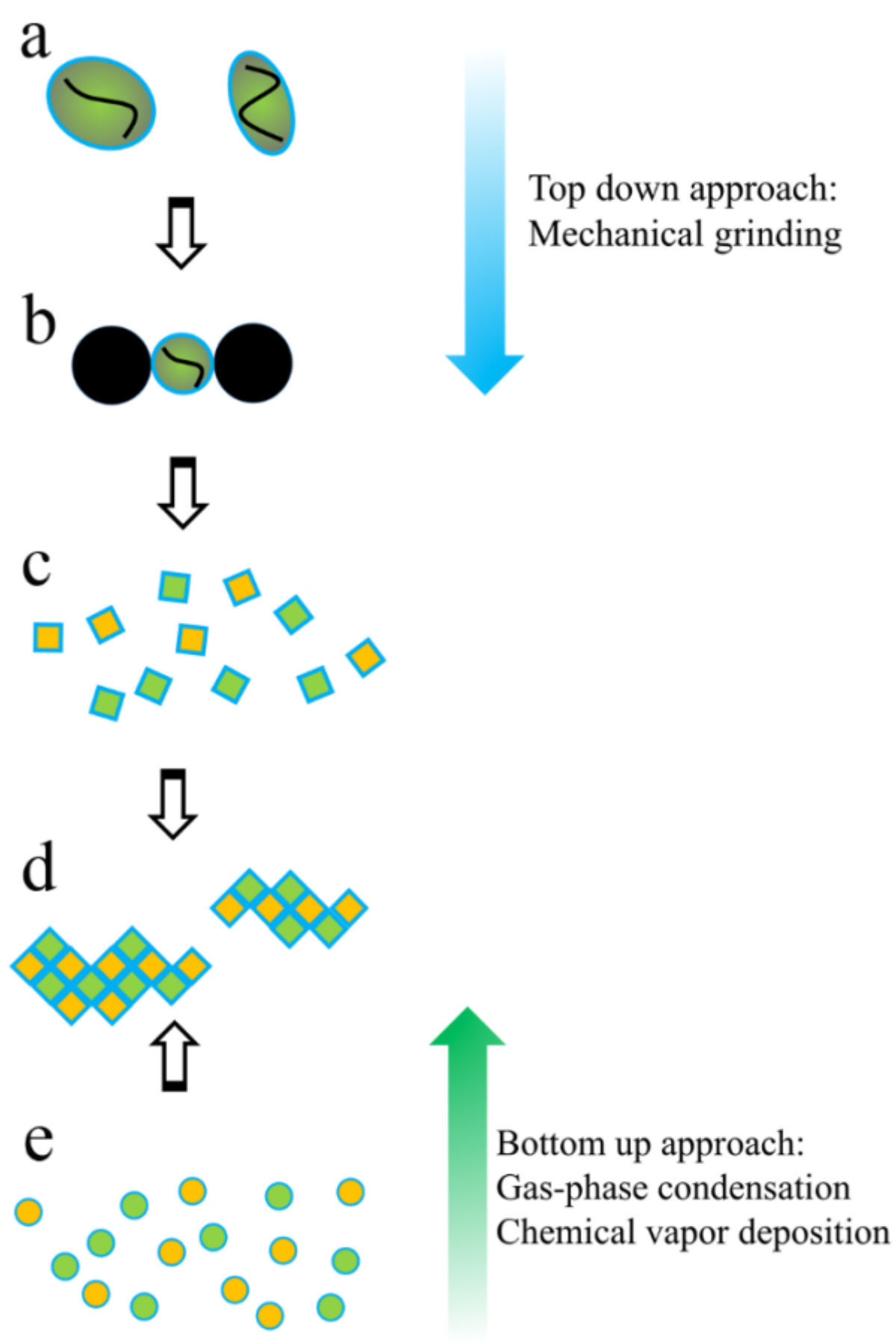

Figure 5. Schematic drawing of a typical top down and bottom up approach for nanoparticle synthesis. (a) Bulk material. (b) Grinding ball. (c) Nanoparticles. (d) Clusters. (e) Atomic level.

### 3.1 Ball Milling Method

Ball milling method developed by John Benjamin [107] in 1970 is used to prepare powders with reduced size. Fecht *et al.* [108] proposed the working mechanism of ball milling method. Schematic drawing of the working principle of ball milling method is shown in Figure 6.

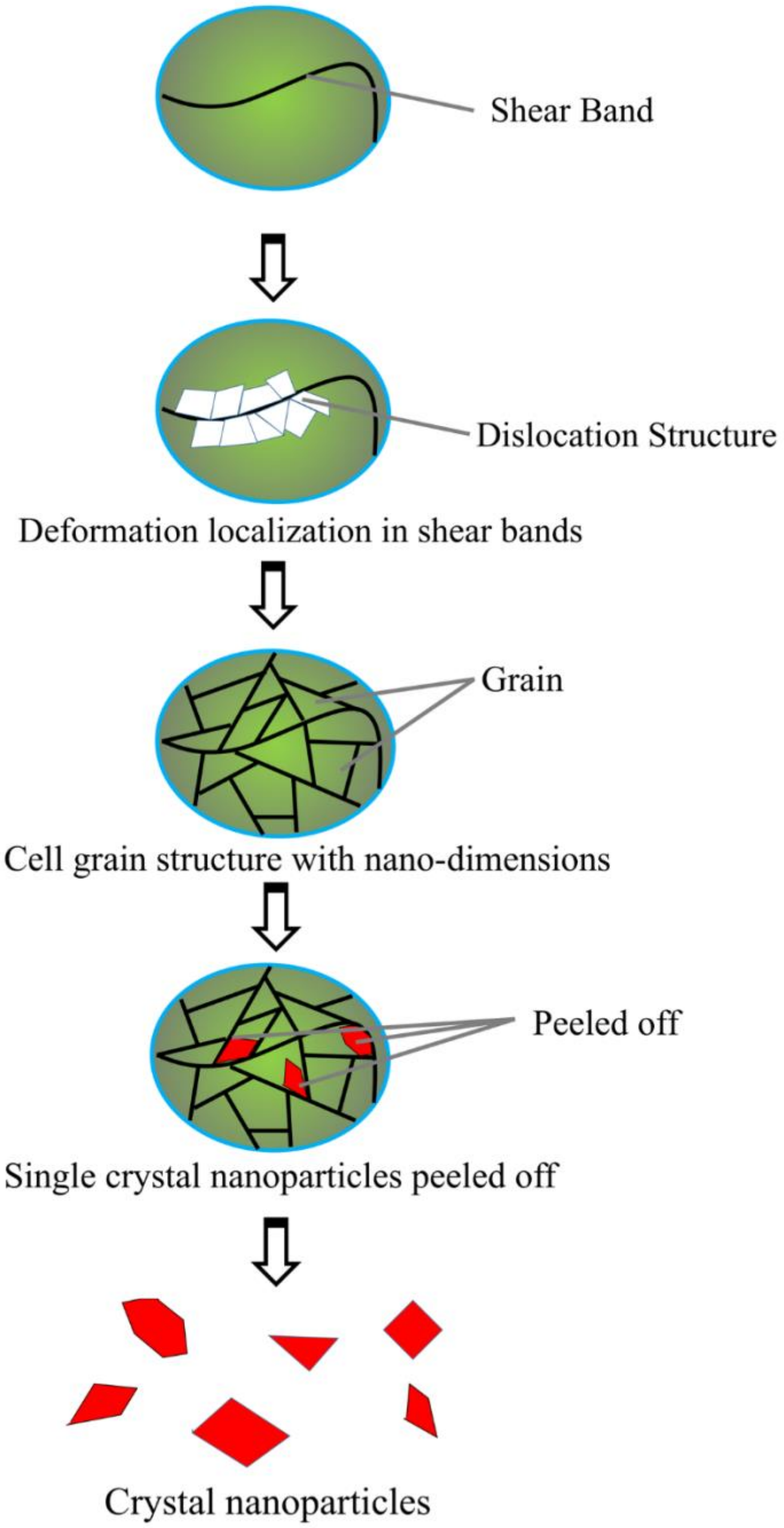

Figure 6. Schematic representation of ball milling mechanism for the formation of crystalline nanoparticles.

There are three stages for ball milling method to prepare nanoparticles. In the first stage, deformations and dislocations are introduced into bulk materials during collisions between balls and bulk materials. The dislocation density keeps increasing with milling time. In the second stage, small grains (nanoscale) are formed due to the accumulation, recombination or rearrangement of dislocations. In the third stage, the grain orientation became random. And then grains at the edge of bulk material are peeled off. Thus, the nanoparticles are obtained from bulk materials. The relationship between grain size of the prepared nanoparticle and the milling time could be estimated by the equation $d = kt^{-2/3}$, where $d$ is the grain size of the nanoparticle, $k$ a constant related with specific system and materials, and $t$ the milling time [109].

However, there is a limitation to the particle size obtained in a specific ball milling system. If the milling time is too long the so-called cold-welding effect, where nanoparticles will be welded together resulting in big particles [110, 111]. When particles have smaller sizes, for examples sub-micrometer, the surface energy of the particles plays a more important role, and particles are more likely to aggregate together to reduce their surface energy. In order to reduce the cold-welding effect, a surfactant is used to lower down the surface energy and reduce the cold-welding effects [112, 113]. Chen *et al.* [110] reported a surfactant-assisted ball milling method to prepare de-agglomeration graphite nanoparticles. The agglomerate size varies from 1 μm to 30 μm and much smaller nanoparticles (< 100 nm) are obtained when using Phosphoric acid dibutyl ester as a surfactant. Even through surfactant-assisted ball milling could produce particles with smaller size, the size distribution is usually very wide. How to control the size distribution of particles prepared by ball milling method is another problem need to be solved. Liu *et al.* [114, 115] proposed an idea to select particles with different size via centrifugal separation with different "settling-down" time of a particle solution. In their experiment, different kinds of nanoparticles, such as Fe, Co, FeCo, SmCo, and NdFeB, are obtained with much narrower size distributions. Nanoparticles with high saturation magnetizations such as Fe, FeCo are focused due to their high magnetic moments. Figure 7 exhibits the TEM images of Fe and $SmCo_5$ nanoparticles prepared by a surfactant-assisted ball milling approach. The Fe nanoparticle size ranges from 4 nm to 6 nm as shown in Figure 7(a) and (b). $SmCo_5$ nanoparticles with size ranging from 3 nm to 13 nm are obtained after 5 h milling, and elongated shape $SmCo_5$ nanoparticles are achieved with even longer milling time of 25 h, as shown in Figure 7(c) and (d). Poudyal *et al.* [116] reported Fe, Co, and FeCo nanoparticles using surfactant-assisted ball milling method. Uniform size (~ 6 nm) particles could be obtained by properly controlling the ball milling parameters and applying size selection process.

For bio-applications, ball milling with a surfactant is used because of reduced cold-welding effect, small particle size and narrow particle size distribution, which makes surfactant-assisted ball milling a suitable way to prepare particles for bio-related applications.

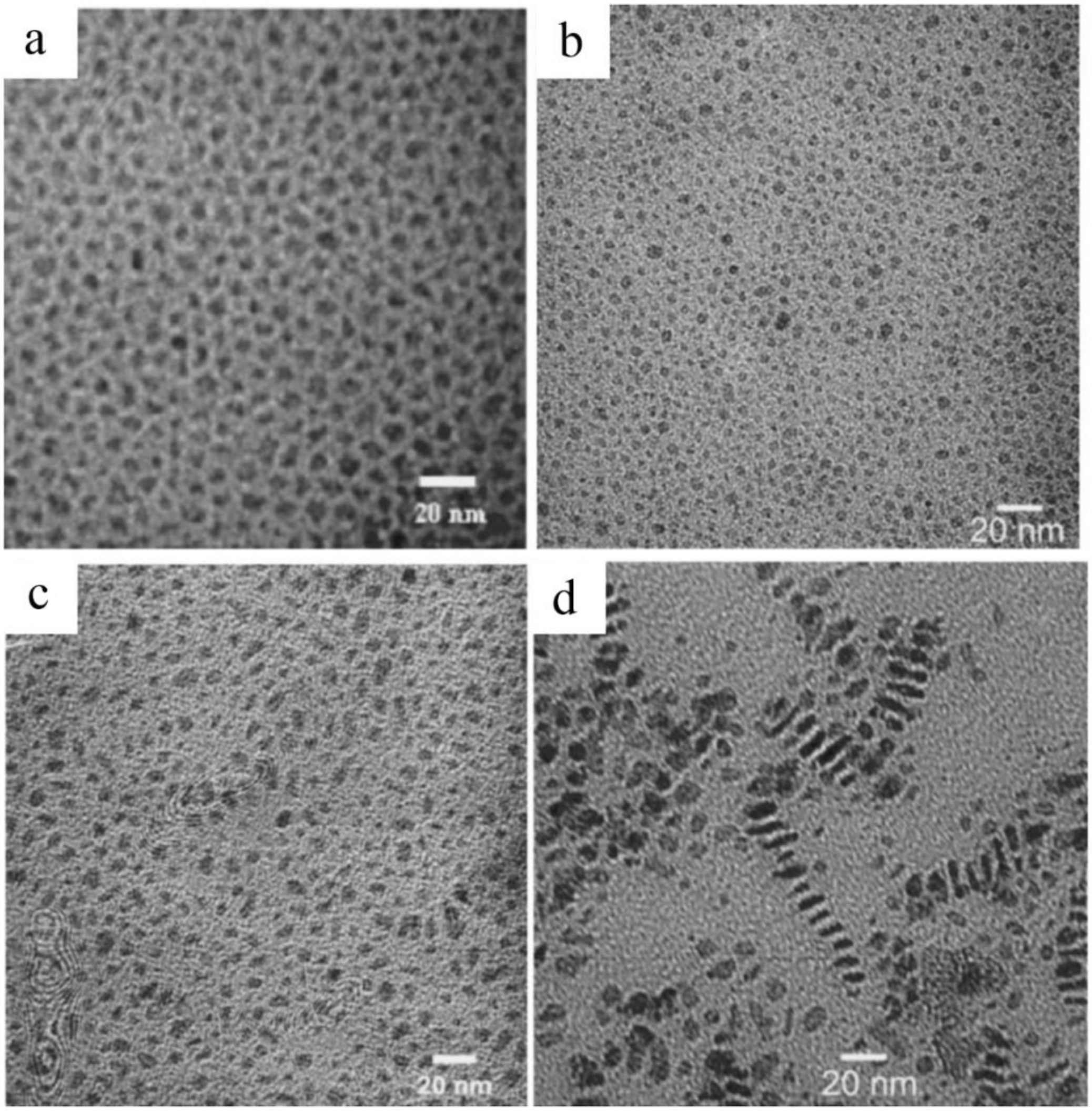

Figure 7. TEM images of the nanoparticles prepared by milling Fe powders for (a) 1 h, and (b) 5h, and by milling $SmCo_5$-based powder for (c) 5 h, and (d) 25 h. (Reprinted with permission from reference [115], copyright 2006 AIP Publishing LLC)

### 3.2 Gas-phase Condensation (GPC) Method

Gas-phase Condensation (GPC) is a bottom-up approach in which atoms (from the sputtering or evaporation) nucleate and grow to form nanoparticles. The particle size and crystallinity are well-controlled by this method while the throughput is generally low compared to ball milling methods. Granqvist and Buhrman [117] firstly used the GPC method to prepare ultrafine particles in 1976. In this GPC system, the atoms are generated from an evaporation source. Then atoms nucleate and grow into nanoparticles in static inert gas. However, both the particle size and crystallinity are out of control. Sattler *et al.* [118] introduced differential pressure and a skimmer into the GPC system, and the size of nanoparticles obtained is under control, but this method is still unable to control the crystallinity of nanoparticles. Later in 1991, a sputtering source is adopted. Thus, more materials are suitable for preparing nanoparticles [119]. Yamamuro *et al.* [120] pointed out the key factor to get monodispersed nanoparticles is to separate nanoparticles' nucleation and growth process into different space regions. In 2006, Wang *et al.* [121] reported that field-controlled plasma heating effect could help control the phase and crystallinity of NPs. For instance, disordered $A_1$ phase and ordered $L_{10}$ phase FePt nanoparticles are successfully prepared using the GPC method by adjusting the plasma heating effects. Moreover, meta-stable phase body-centered tetragonal Fe nanoparticles are also successfully made by this method [122].

We will focus on GPC system with sputtering sources in this section. In the GPC system, atoms are kicked off from a target forming atom vapor near the target surface. High sputtering pressure (several hundred mTorr) is applied to form nanoparticles instead of thin films since the mean free path of atoms is much smaller at high sputtering pressure (several hundred mTorr). In this case, these sputtered atoms will collide with argon atom, and energy will transfer from atoms to argon atoms. Thus, the temperature of atoms is cooling down and nucleation and growth will happen when the temperature is low enough. The size of the nanoparticle is dependent on the sputtering current, magnetic field intensity at the target surface, sputtering pressure and gas flow rate. Due to magnetron sputter sources, many kinds of particles have been successfully prepared, such as high magnetic moment MNPs (Fe, Co, FeCo), core-shell nanocomposite, and MNPs with the tunable magnetic property.

First, well-crystallized high moment FeCo nanoparticles are successfully synthesized by a GPC method. The average size of FeCo nanoparticle is around 12 nm showing superparamagnetic properties, which is suitable for bio-related applications such as GMR based biosensors [123, 124]. As shown in Figure 8, both spherical and cubic shape FeCo nanoparticles could be made using GPC method by adjusting the plasma heating effects. Some other high moment nanoparticles such as Fe and Co are also successfully synthesized using a GPC method.

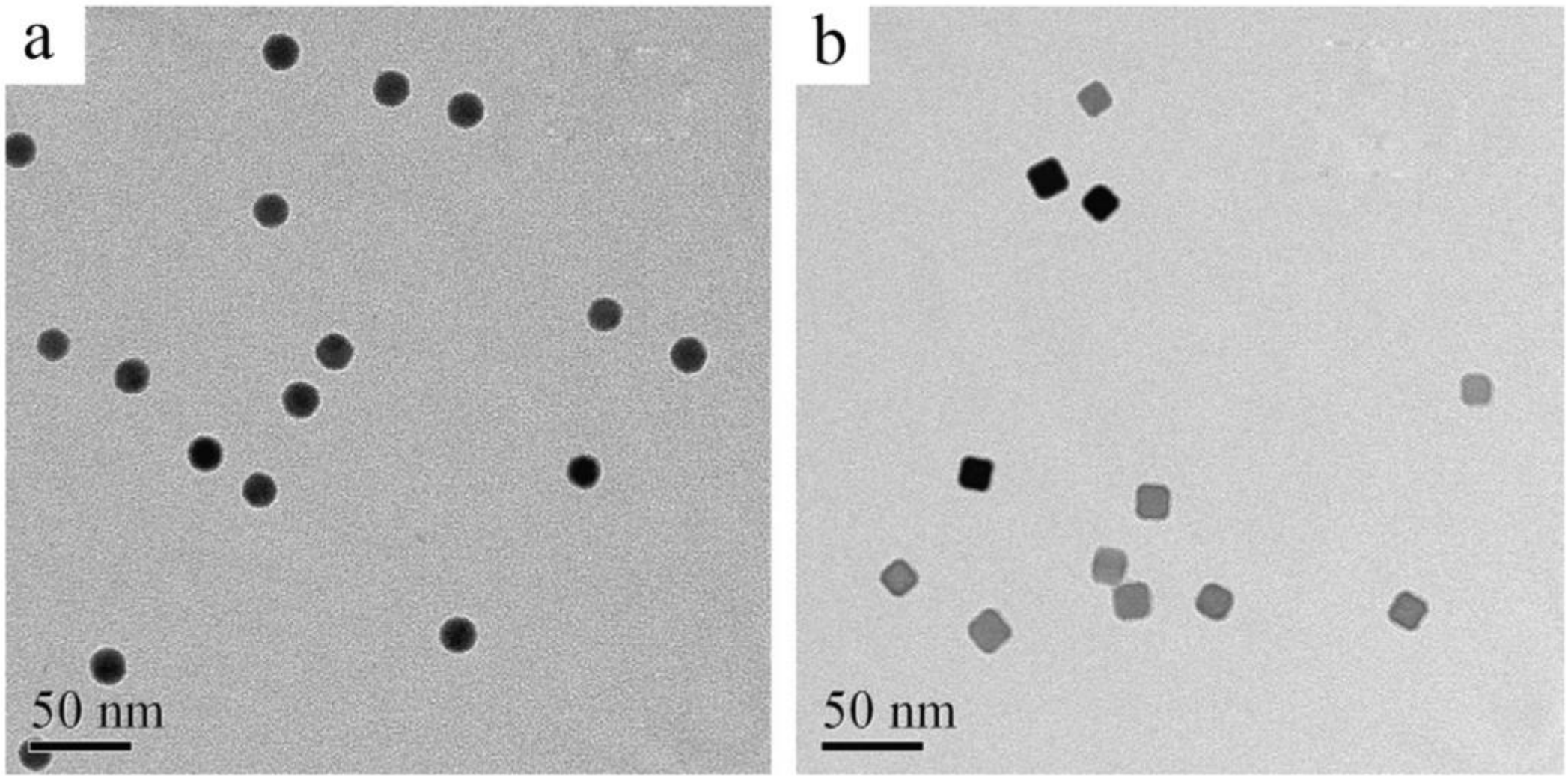

Figure 8. TEM bright field images of FeCo nanoparticles with different shapes, (a) spherical and (b) cubic. (Reprinted with permission from reference [125] , copyright 2007 Elsevier)

Second, to make nanoparticles easier for following surface chemical modification, core-shell nanoparticle structure is a good candidate. Gold is good for the surface modification, making it a good candidate for shell material for high moment magnetic nanoparticles. To make a gold shell for magnetic nanoparticles, atom diffusion at the nanoscale should be well-understood. There are two effects competing during shell formation: diffusion via concentration gradient and surface segregation. The first one is to make atoms distribution uniform and the other one is opposite. Therefore, surface segregation should be a major effect in obtaining core-shell nanoparticles [126]. Figure 9 shows the TEM images of FeCo nanoparticles with Au shell prepared by the sputtering-based GPC method. Core-shell nanoparticles could also be prepared by multiple ion cluster source (MICS), in which three independent magnetron sources replace the single source as discussed above [127]. In MICS setup, one source could be used as core NPs synthesize and others for shell materials. In this case, some core@shell, core@shell@shell structures are prepared [128]. As shown in Figure 10, nanoparticles with Co/Ag/Au core@shell@shell structure are prepared, demonstrated by scanning TEM (STEM) image and energy dispersive X-ray spectroscopy (EDS) line scan and electron energy loss spectroscopy (EELS) element mapping of these nanoparticles.

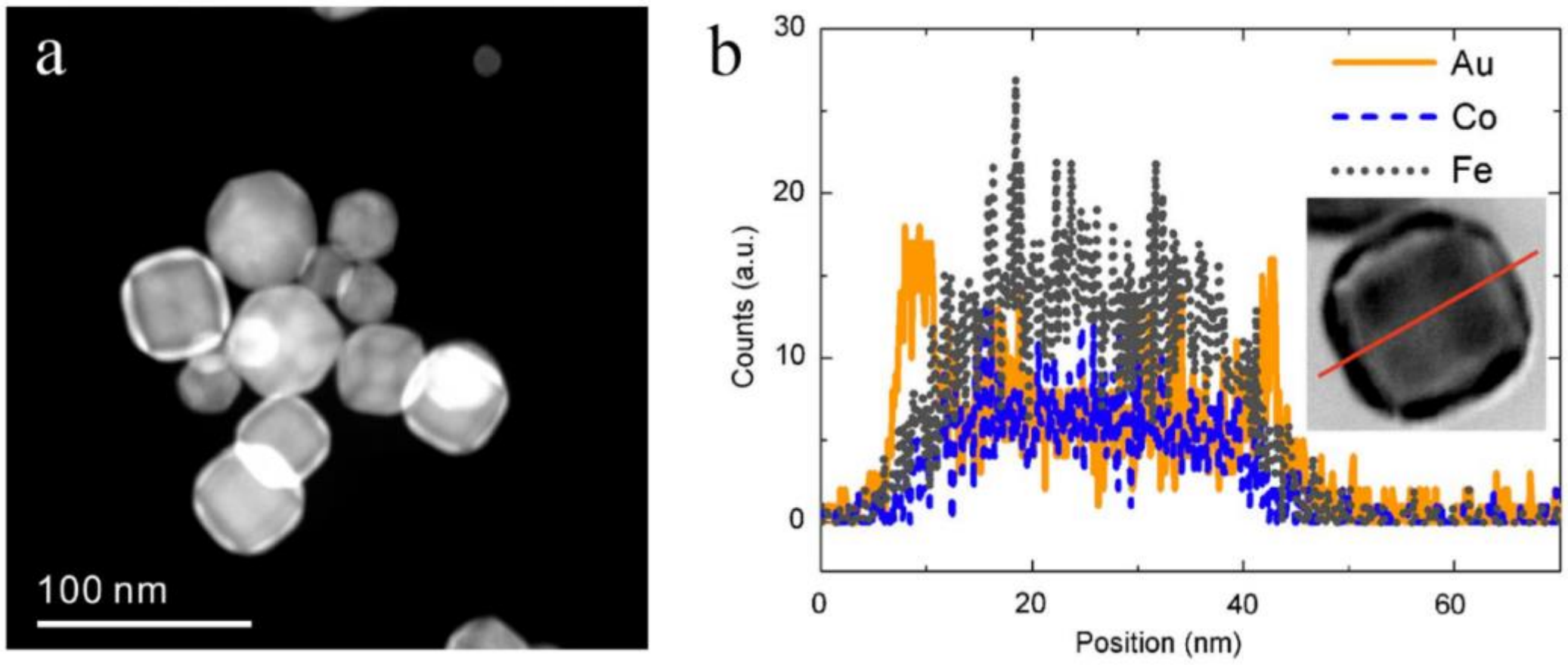

Figure 9. Morphology and composition distribution of FeCo–Au nanocrystals. (a) High angle annular dark field (HAADF) image showing clear core-shell structure due to the contrast between the different materials of core and shell. (b) Composition distribution of a cross section of a single FeCo–Au nanocrystal, acquired by EDS line scan. The nanocrystal is shown in the inset. The line indicates the path of the electron beam. (Reprinted with permission from reference [129], copyright 2007 AIP Publishing LLC)

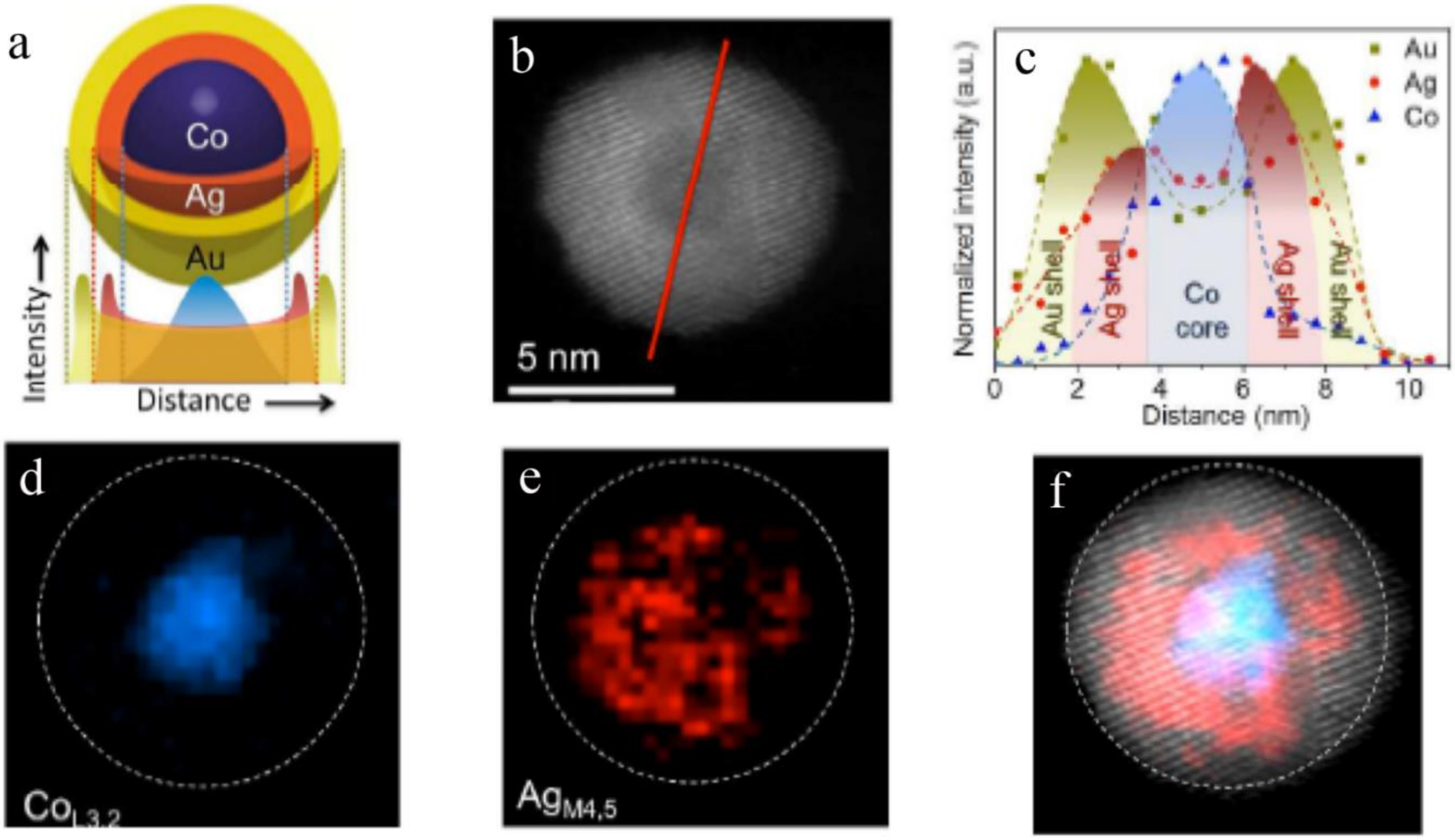

Figure 10. Core@shell@shell Co@Ag@Au nanoparticles. a) Representation of the complex Co@Ag@Au structure together with the expected EDS intensity profiles. b) Cs-corrected STEM representative image of a Co@Ag@Au NP. c) EDS line scan performed at the Co, Ag and Au, along the line depicted in Figure 4b. d) EELS compositional analysis for the Co $L_{3,2}$ edge. The dashed line represents the outer limit of the NP. e) EELS map for the Ag $M_{4,5}$ edge. f) STEM image together with the corresponding Co and Ag EELS concentration maps superimposed. (Reprinted with permission from reference [128], copyright 2014 The Royal Society of Chemistry)

### 3.3 Other Nanoparticle Fabrication Approaches

Besides ball milling method and gas-phase condensation system, there are some other approaches to prepare nanoparticles such as wet chemical way [130, 131], chemical vapor deposition [132, 133], thermal decomposition [134, 135], etc. Figure 11 shows the SEM and TEM images of FeCo nanocubes prepared by a wet chemical method. The size of FeCo nanocubes ranges from 60 nm to 270 nm. The morphology and dimensions of the FeCo nanotube can be adjusted by controlling the concentration of cyclohexane and PEG-400, the reaction time, and the molar ration of $Fe^{2+}$ and $Co^{2+}$ in the reaction solutions. The saturate magnetization of 68 ± 6 nm FeCo nanocubes is 211.9 emu/g.

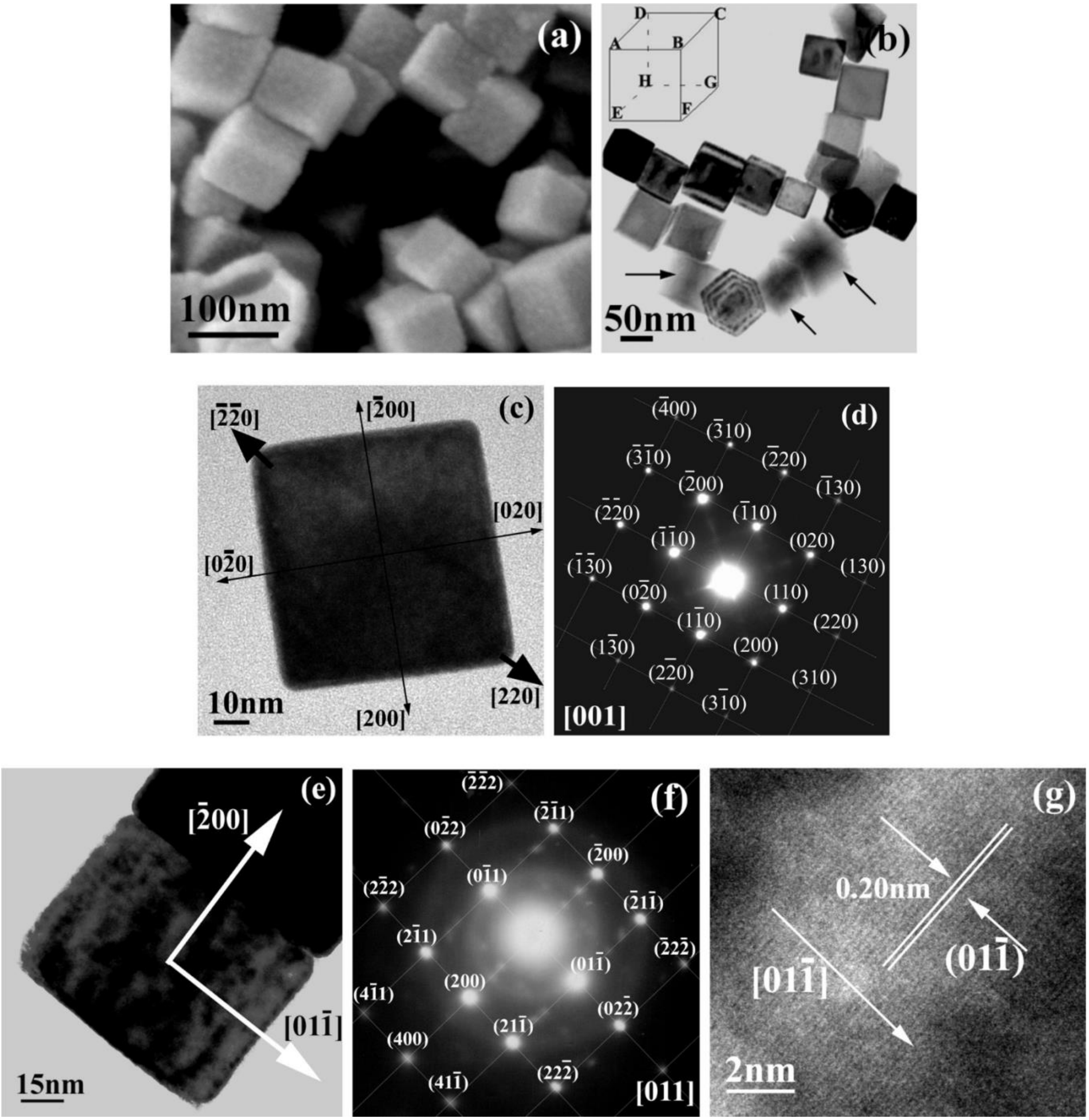

Figure 11. (a) SEM and (b) TEM micrographs of FeCo nanocubes, obtained by reaction of 0.1 M $Fe^{2+}$ and $Co^{2+}$ with hydrazine for 30 min in the presences of 2.8 M PEG-400 and 0.14 M cyclohexane. The inset is an illustration of the nanocube. (c) TEM image and (d) SAED pattern of a single FeCo alloy nanocube oriented along [001]. (e) TEM image, (f) SAED pattern, and (g) HRTEM image of a FeCo alloy nanocube oriented along [110]. (Reprinted with permission from reference [130], copyright 2008 American Chemical Society)

# 4 Surface Coating Strategies

The biocompatibility and chemical stability of MNPs can be enhanced by conjugating chemical compounds such as polyethylene glycol (PEG), chitosan, lipid, proteins, etc. [136]. In most biological applications, the solubility and chemical stability of the MNPs should be well controlled in different environments both in vivo and in vitro. It is also necessary to prevent MNPs from aggregation and precipitation while maintaining their biocompatibility and chemical stability [136]. To this end, surface coating strategies are needed to facilitate the application of MNPs in nanomedicine. In this section, both organic and inorganic coating strategies with regard to different areas of applications will be reviewed.

## 4.1 Organic Coating

MNPs synthesized through organic solutions are monodisperse, single crystalline with the high magnetic moment. But since these particles often turn out to be hydrophobic, additional surface modification techniques are needed [137]. In general, the organic coating techniques on MNPs can be divided into covalent and absorption processes. As is shown in Figure 12, the covalent coating can be further divided into grafting-to, grafting-from, and grafting-through techniques [138].

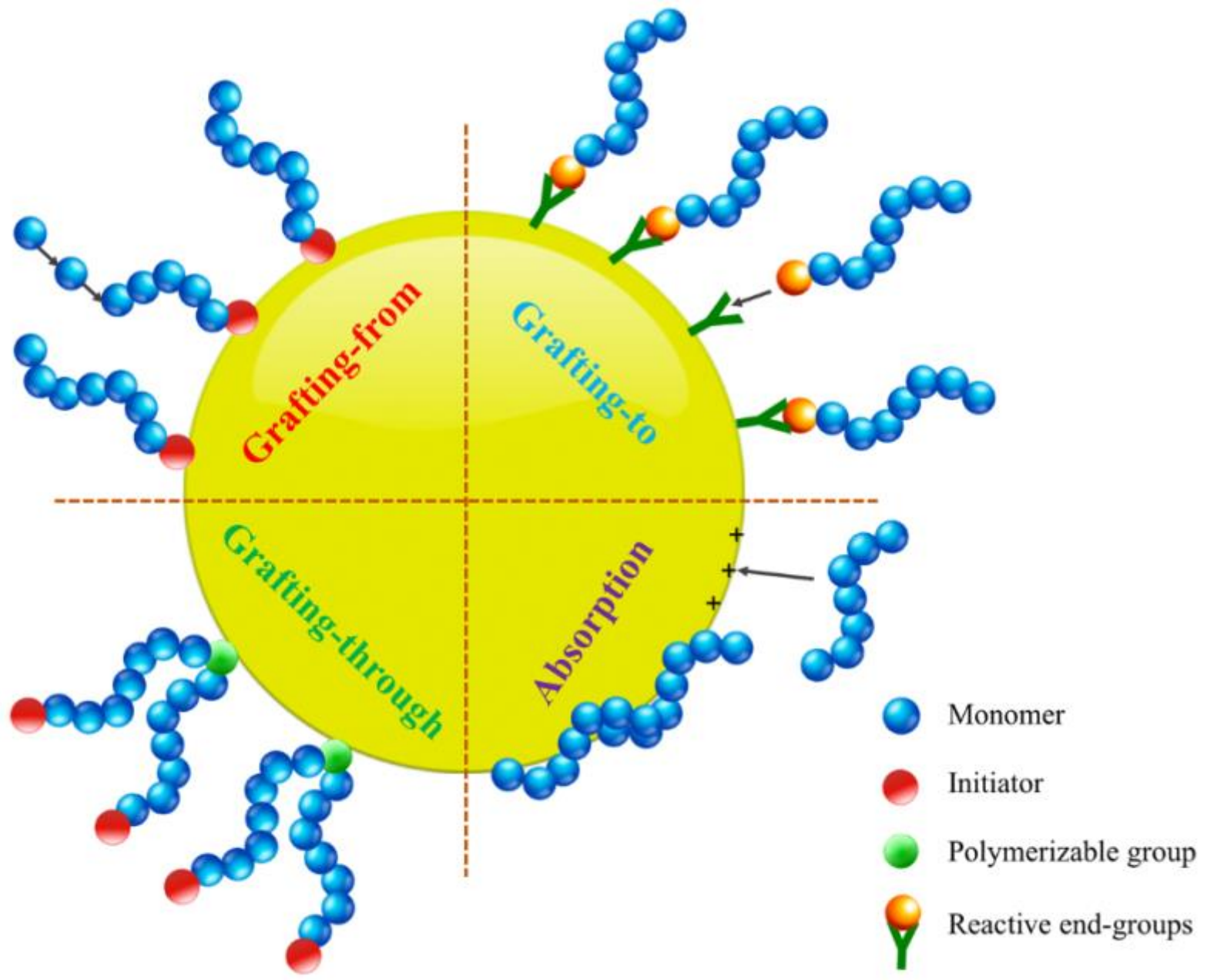

Figure 12. Schematic representation of different coating techniques for MNPs.

By reacting with hydrophilic molecules, the hydrophobic surfactants on the nanoparticles can be replaced with hydrophilic ligands. Polyethylene glycol (PEG) is the most commonly used polymer for such ligand exchange process due to its biocompatibility and also the ability to reduce any non-specific reaction between MNPs and proteins [139]. For example, Xie *et al.* [140] synthesized monodispersed 9 nm $Fe_3O_4$ nanoparticles and functionalized them with dopamine (DPA) terminated PEG. The coating process is shown in Figure 13. Since DPA moiety has higher affinity to the $Fe_3O_4$ surface, it was first linked to PEG and was then used to replace the oleate and oleyamine on the particle surface. It was found that not only did these MNPs have higher stability and no agglomeration in water and physiological environment, they were also undetectable by the body immune system, which made them promising candidates for drug delivery applications. Besides PEG, other polymers such as dendrimers [141], polyethylene oxides [142] and dextrans [143] have also been used in the ligand exchange. In addition, small molecules that have high affinity to the MNP surface, such as the previously mentioned DPA, can also be directly used in the ligand exchange [144, 145].

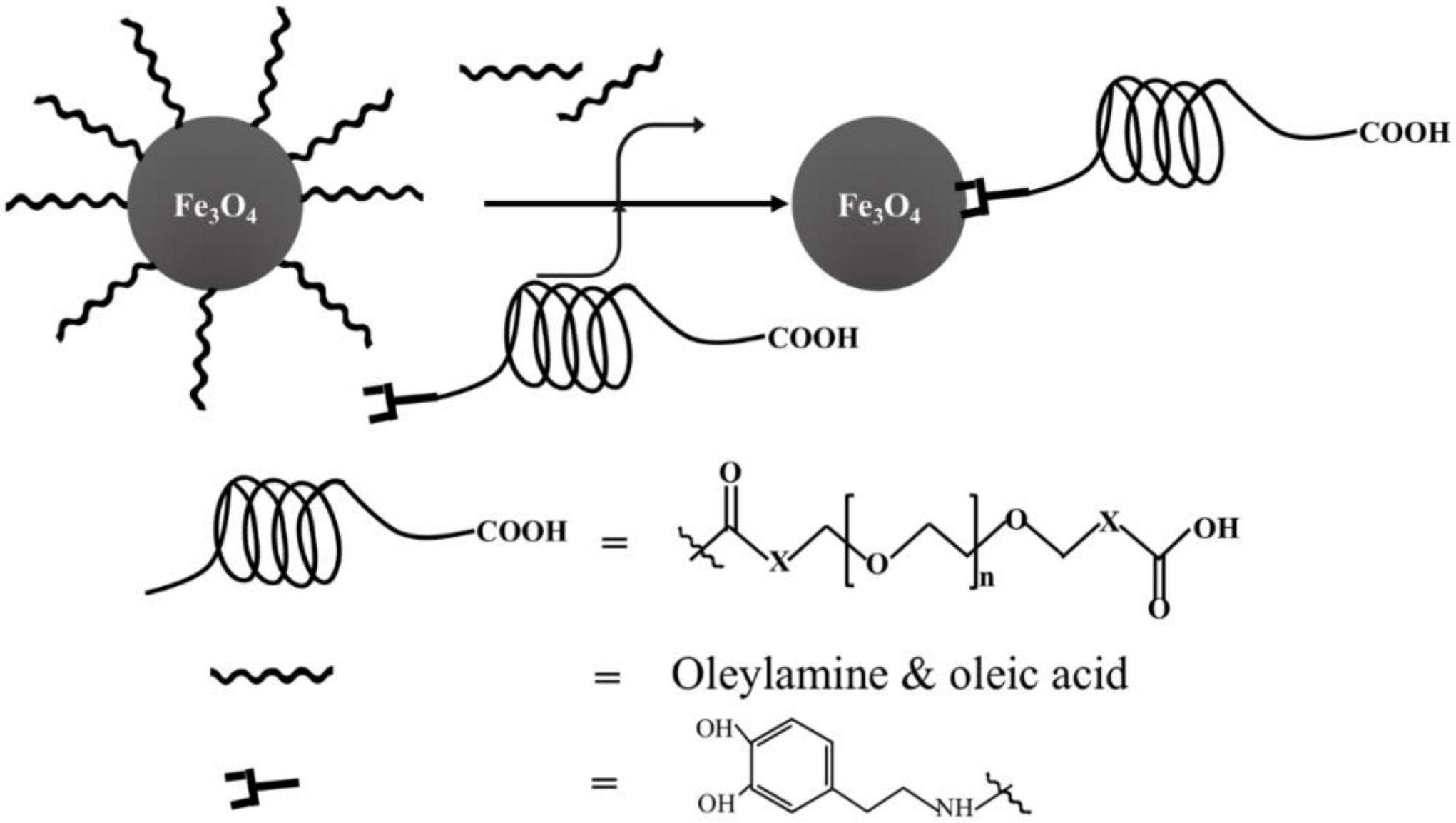

Figure 13. Surface modification of $Fe_3O_4$ nanoparticles via DPA-PEG-COOH. X=$CH_2NHCOCH_2CH_2$.

Although PEG is commonly used for the requirement of minimizing non-specific protein absorption, it has a major drawback, which is the tendency of oxidative degradation. To solve this problem, polyzwitterions, which are made from zwitterionic repeat units without net charge, were developed [138]. Since zwitterionic moieties are the main component of cell membranes and exhibit long circulation time, which can largely reduce the cytotoxic side effects. Several approaches are available for the functionalization of polyzwitterions. For example, von der Lühe *et al.* [146] successfully coated polydehydroalanine (PDha) on the surface of MNPs. The functionalization can also be accomplished by simply inducing polyzwitterions during the synthesis of MNPs.

In addition to surface coating, MNPs can also be encapsulated in a shell to increase their biocompatibility and hydrophilicity. Cheng *et al.* [147] embedded MNPs in a copolymer matrix, i.e., poly(D,L-lactide–co–glycolide)–block–poly(ethylene glycol) (PLGA-b-PEG-COOH) and managed to tune the size of the MNPs by altering solvent type, polymer concentration, and solvent : water ratio. They were the first group to report a linear correlation between polymer concentrations and the size of the resulting MNPs, which makes it possible to precisely control the size of the MNPs to fulfill different requirements for drug delivery in various organs. Other copolymers, such as polystyrene-co-poly-(acrylic acid) (PS-co-PAA) [148] and poly(D,L-lactide)–block–poly(ethylene glycol) (PLA–b–PEG) [149] are also used in similar applications.

## 4.2 Inorganic Coating

Silica is one of the most widely used inorganic coating materials, which is biocompatible and can provide non-aggregated and stable suspensions [150]. A silica outer shell also allows for the subsequent functionalization of alkoxysilanes. Chen *et al.* [151] demonstrated the nontoxicity of FePt NPs with a 17 nm thick silica shell and proved that FePt NPs can be internalized by tumor cells. Furthermore, it was found that these particles were strong $T_2$ agents and thus had great potential for the design of diagnostic and therapeutic agents for drug delivery and hyperthermic tumor ablation. The capability of cellular imaging and *in vivo* MRI applications was also demonstrated.

To obtain a better control of the magnetic properties of the silica-coated MNPs, especially to satisfy the temperature requirement for hyperthermia treatments, the influence of silica coating on the magnetization and the Currie temperature of the MNPs were studied. It was shown that at room temperature, the magnetization can be decreased by 32% by silica coatings under a magnetic field of 1 kOe due to the presence of diamagnetic shells [152]. The Currie temperature can also be reduced by 7% due to the interaction between silica and the MNP surface. Thus, a trade-off between the biological stability and magnetic properties should be considered for silica coated MNPs.

Compared to organic coatings, inorganic coatings are more frequently used in hyperthermia treatments. Since metals are conducting materials, metallic coatings allow for inductive heating under an AC magnetic field. Due to its easy conjugation with many biomolecules such as DNAs and proteins, gold has become the optimal metallic coating for MNPs [153]. While gold coatings can simplify subsequent bio-functionalization processes and protect the internal MNPs from oxidation, they inevitably induce changes in the magnetic properties of the resulting NPs. Presa *et al.* [154] synthesized FePt nanoparticles from high-temperature solution phase and tried to optimize the number of gold coatings on the MNP surface. It was shown that at low temperature, the coercivity of the FePt-Au NPs decreases about 3 times and the blocking temperatures also reduce to the half compared to uncoated NPs, which was attributed to the reduction of surface anisotropy due to the interaction between gold atoms and the FePt NPs. Yano *et al.* [154] also did a similar study on this issue and pointed out the direct relationship between coercive field and magnetic anisotropy, but the mechanism underlying in the gold-FePt interaction is still not clear.

# 5 MNPs for Diagnosis

Early detection of diseases allows for therapeutic intervention in the early stages, which is the key to successful treatment. With the rapid developments in nanotechnology, optical and electrochemical sensors have been applied as high sensitivity diagnosis platforms. However, considering their susceptibility to interferences from unprocessed biological samples, the optical and electrochemical sensors require complicated sample pretreatments [155]. On the other hand, magnetic nanomaterial-based sensing systems are attracting much attention for the immunoassay applications since biological samples exhibit negligible magnetic susceptibility, which makes very high signal-to-noise ratio (SNR) magnetic sensing possible even in a minimally processed biofluidic sample.

## 5.1 Magnetoresistive (MR) Sensors

### *5.1.1 Giant Magnetoresistive (GMR) Biosensors*

Since its discovery in 1988, GMR sensors have been widely used in hard disk drives [156, 157], position sensing [158], electrical current measurement [159, 160], as well as biomarker detection [161, 162 163, 164]. GMR effect is observed in structures with alternating ferromagnetic and nonmagnetic metal layers (see Figure 14(a)). When exposed to the external field, the magnetization orientation of the "free" ferromagnetic layer with respect to the "pinned" ferromagnetic layer will change, which will result in the change of the resistance of the whole device [165, 166]. The ability of a GMR sensor to respond to the external field is characterized by the GMR ratio, which can be expressed as:

$$GMR = \frac{R_{AP} - R_P}{R_P} \times 100\% \quad (19).$$

The first GMR based sensing system for biomarker detection was performed by Baselt *et al.* [167] The Bead Array Counter (BARC), which includes multilayer GMR sensors and magnetic microbead tags, was developed and exhibited great potential in the measurement of intermolecular forces during biomolecular recognition processes. In the past 20 years, lots of research has been carried out to optimize the structure of the magnetic immunoassay and the performance of the GMR detectors. The most widely used magnetic immunoassay is built up based on antibody-antigen reactions. For example, in the GMR based probe station system developed by Wang *et al.*,[168, 169] the MNPs were functionalized with streptavidin, which can bind to the biotinylated detection antibody. During the detection process, the analyte was captured by the capture antibody functionalized on the GMR sensor surface. Then, the detection antibody was added and can only bind to the sites with the analyte. The functionalized MNPs were subsequently attached to the detection antibody as the final layer. Under the external field, the MNPs were magnetized, whose dipolar field (also known as the stray field) can be sensed by the GMR sensors underneath [170] (see Figure 14(c)). To realize on-site diagnosis, several handheld systems were also developed [171, 172]. With the size of a snack container and a user-friendly interface, the system can be used by non-technicians with minimum expertise (see Figure 14(e)(f)). Apart from antibody-antigen reactions, aptamers can also be used in the immunoassay (see Figure 14(d)) [173].

The MNPs used in the immunoassay are often embedded in a polymeric matrix (also known as multi-core MNPs) and their diameters can vary from tens of nanometers to hundreds of nanometers [174-177]. Compared to other biomarker tags, MNPs exhibit superior performance due to low background noise, low possibilities of aggregation due to superparamagnetism, and high biological compatibility [178]. To date, the detection of various biological targets has been demonstrated, such as DNAs [173], viruses [171, 179] and food pathogens [180, 181]. Chugh *et. al* [182-184] has demonstrated that GMR in its IC form has also been quite active in non-invasive determination of primary healthcare parameters such heart rate, respiration rate and blood pressure.

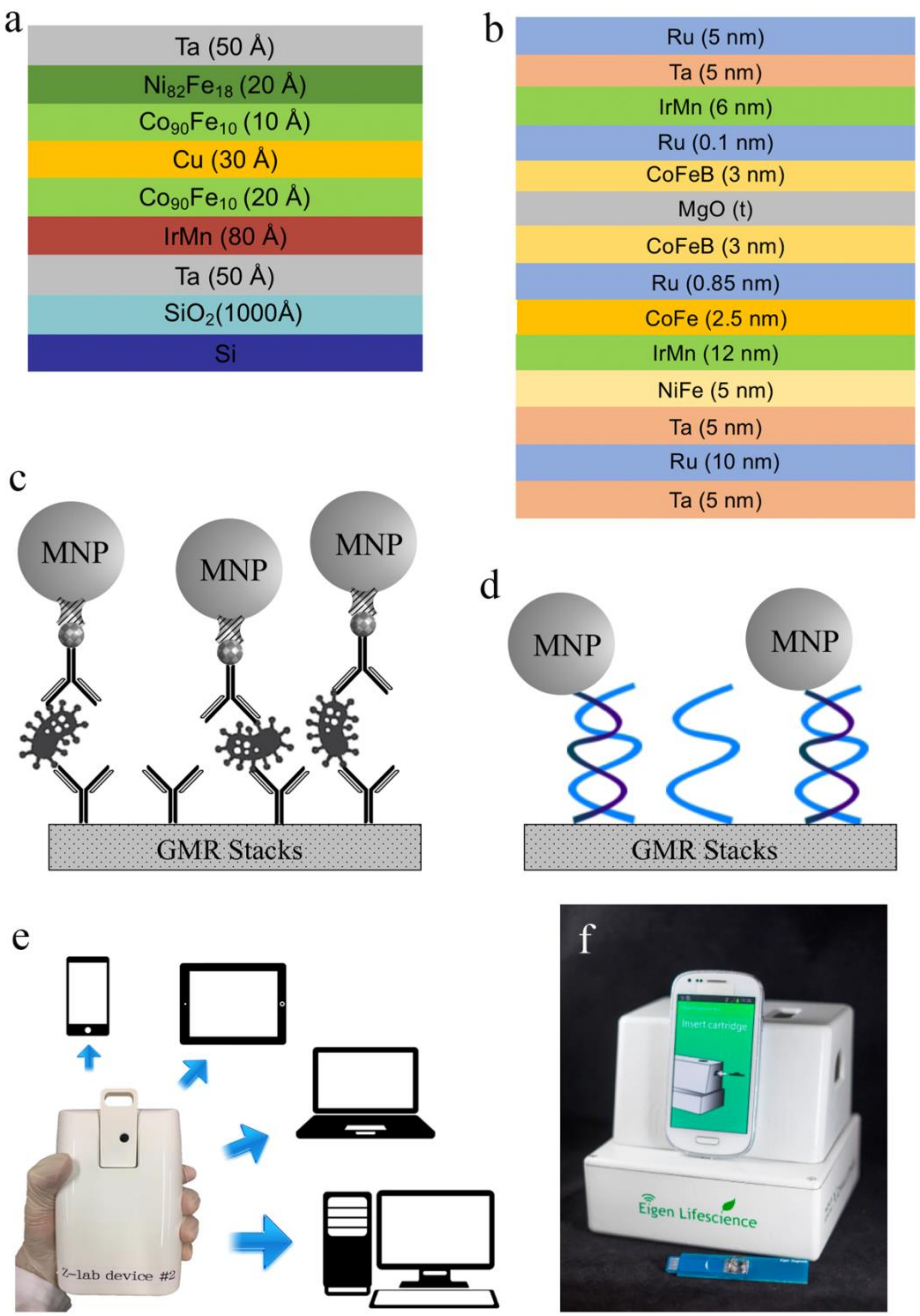

Figure 14. (a) An example for GMR spin valve stacks. (b) An example for MTJ stacks. (c)Magnetic immunoassay on GMR sensor surface. (d) GMR detection based on aptamers. (e) and (f) are GMR-based handheld systems developed by two different groups. ((e) reprinted with permission from reference [171], copyright 2017 American Chemical Society, (f) reprinted with permission from reference [172] , copyright 2016 Elsevier)

However, a major problem associated with GMR-SV sensors is that they exhibit Barkhausen noise due to the formation and depletion of the magnetic domains during the spin direction reversal of the sensing layer. This causes changes in resistance which may occur in the absence of the MNPs, and when used for the detection of MNP-labels/biomarkers, it can result in false signal detection. Nevertheless, this problem can be resolved if necessary by applying a strong pre-magnetizing field is applied (e.g., by integrated microconductors) to reduce the order the internal magnetic domains prior to the sample run [185].

### *5.1.2 Magnetic Tunnel Junction (MTJ) Biosensors*

In addition to GMR sensors, MTJ sensors are also promising candidates as the detectors for the magnetic immunoassays. The basic structure of a MTJ consists of a thin insulating tunnel barrier sandwiched between two ferromagnetic layers (Figure 14(b)). The tunneling magnetoresistance (TMR) ratio is defined in the same way as the GMR ratio. While the GMR ratio of the spin valve sensors commonly used in the bio-detection is less than 20% [168], the TMR ratio of the MTJ sensors with MgO tunnel barriers can be over 200% at room temperature [186]. MNPs and the immunoassays are integrated with MTJ sensors in the same way as with GMR detectors. Sharma *et al.* [187] have developed a sensing system based on MTJ sensors for the detection of pathogenic DNA. Firstly, the probe oligonucleotides complementary to the target DNA strands are immobilized on the sensor surface. After hybridization between probe DNA and target DNA, the streptavidin-coated MNPs were added. The real-time signal change of the injection, binding and washing steps of the MNPs can be read out from the MTJ sensor arrays. The detection limit for *Listeria* DNA can be as low as 1 nM.

Cardoso *et al.* [188] studied the difference of MTJ sensors and spin valve (SV) sensors in the detection of 130 nm MNPs. It was found that MTJ sensors had a higher signal-to-noise ratio (SNR) than SV sensors but will reach a limit at a current of 900 μA. The SNR of SV sensors, however, will increase continuously with the applied current and will only be limited by the heat generation. In the detection of MNPs, SV sensors also exhibited higher signal level and sensitivity despite the high TMR ratio of the MTJ sensors, which can be attributed to the increased distance between MNPs and sensor surface due to the top electrodes of MTJ sensors. Despite greater linearity and sensitivity of TMR sensors over SV, a major problem for TMR is shot noise that arises from the discontinuity in the electron transport paths is present in MTJs unlike that in spin valves and GMR multilayers. This is due to the existence of an insulating barrier in MTJs; the conduction medium is discontinuous for MTJs.

Another probable source of error in both SV and MTJ sensors can be from the sample itself in cases where there are MNPs from clusters, may be because of the presence of surface charges resulting in an erroneous detection [185]. Such clustering can be avoided, with suitable surface modification and sheathing of the magnetic material as discussed previously in Section 4. Other possible sources of errors in MR sensors are the thermal noise (which are non-magnetic in origin but has a direct effect on the resistance of the sensor) and stray field from MNPs. New, exciting applications of these sensors could be in MCG and magnetic encephalography provided there field of operation can be shifted to elevated frequencies, giving them better noise level, and multiple calibration steps are carried out before use.

## 5.2 Micro-Hall (μ Hall) Sensors

Since Hall voltage is proportional to the external field, Hall sensors can provide a linear response to the dipolar/stray field from MNPs and are only sensitive to its perpendicular component. Thus, they can map out the trajectory of moving magnetic particles and have been used in many areas such as drug delivery, medical imaging, and biomarker detection [189]. In 2002, Besse *et al.* [190] fabricated a highly sensitive silicon Hall sensor with the ability to detect one single magnetic microbead of 2.8 μm by standard complementary metal-oxide-semiconductor (CMOS) technology. Later, devices based on InAs quantum well (QW) semiconductor heterostructures were developed. Phase-sensitive detection was employed and demonstrated even better SNR for micrometer and nanometer-sized magnetic particles [191]. Michele *et al.* [192] also performed single-particle detection using InSb double Hall cross with an area of 1 $\mu m^2$. By sweeping DC field and taking the normalized difference between the in-phase signals, the susceptibility curve for the target magnetic particle can be measured.

Aledealat *et al.* [193] was the first group to integrate microfluidic channel with micro-Hall sensors (see Figure 15). With a cross-shaped InAs quantum well micro-Hall sensor, the real-time signal from beads moving within and around the sensing area in the microfluid channel can be detected and distinguished by polarities. The first attempt to use micro-Hall sensors in biological detection was made by Issadore *et al.* [194], in their work, the MNPs were bound to the bacterial cell wall and acted as the magnetic tags. The stream with target cells was flowed through the hall sensor surface in a microfluidic channel and was confined in the vertical direction. It was shown that the sensing system can reliably distinguish Gram-positive from Gram-negative species and requires much smaller sample volume (1 μL) compared to flow cytometry. The detection limit of the system was ~10 bacteria, which was comparable to that of culture tests, but the assay time was 50 times faster. Sandhu *et.al* [195] had studied some practical sensors based on Hall Effect for biomedical instrumentation.

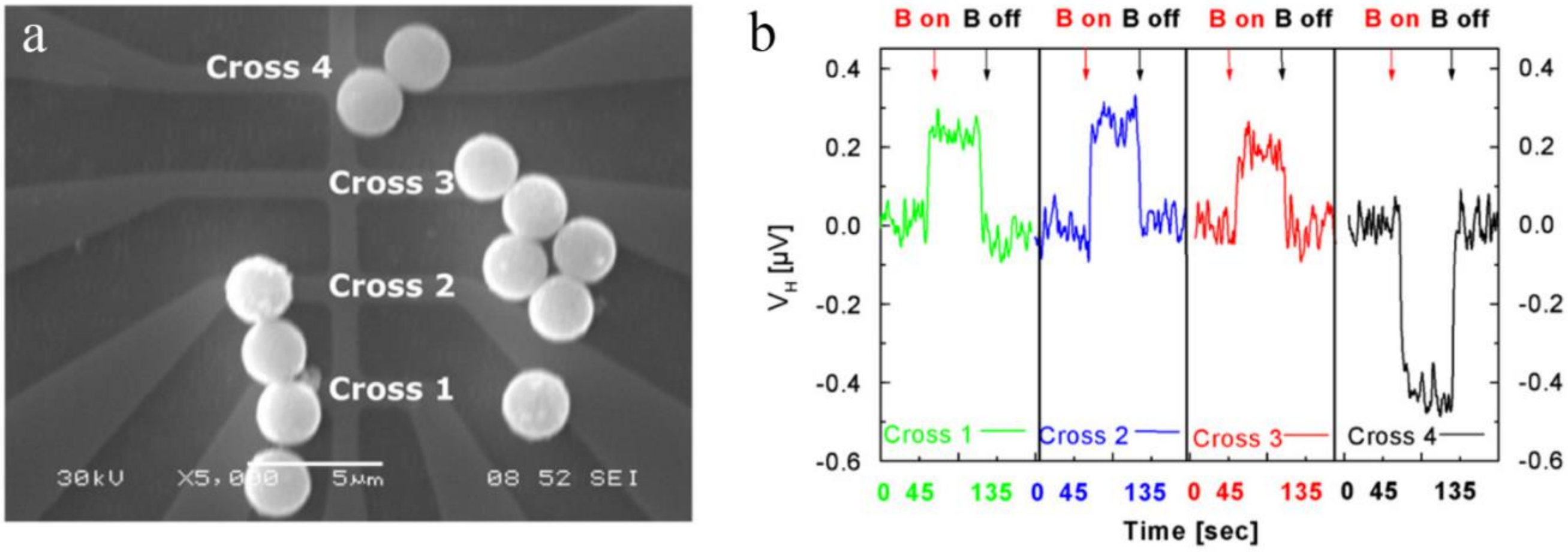

Figure 15. Schematic illustration of the microfluid channel integrated with micro-Hall sensor. (a) SEM image of the central region of an InAs micro-Hall sensor and immobilized SPM beads. (b) $V_H$ as a function of time for crosses (1)–(4). The increase in $V_H$ for crosses (1)–(3) and its drop for cross (4) as B was applied agrees with the expected signals based on a dipolar stray field representation of SPM beads. (Reprinted with permission from reference [193], copyright 2010 Elsevier)

## 5.3 NMR-based Diagnostics

NMR-based diagnostics exploits MNPs as proximity sensors (contrast agents), which modulate the $T_2$ of water molecules adjacent to the target-MNP aggregates. When MNPs specifically bind to their target molecules through affinity ligands, they form magnetic clusters which lead to a faster decay of NMR signal or shorter transverse relaxation time $T_2$ of the surrounding water protons. The NMR-based biosensing technology collects signals directly from the whole volume of sample (volume-based biosensor), which effectively shorten the immunoassay time than that of surface-based sensors such as GMR, MTJ, and µHall sensors. This class of sensors (volume-based biosensors) are more flexible, smaller in size, and suitable for on-site diagnosis.

### *5.3.1 Magnetic Relaxation Switching (MRSw) Assay*

As is mentioned in section 2.5, the transverse relaxivity $r_2$ of MNPs are greater than longitudinal relaxivity $r_1$, thus MNPs are mainly used as $T_2$-modulating agents for NMR related applications. In MRSw-based nanosensors, the change of $T_2$ mainly comes from the aggregation degrees of MPNs in the presence of target analytes. To date, there are two types of detection mechanisms that have been reported so far [196-200], namely, $T_2$ decreases with the aggregation of MNPs in type I system (see Figure 16(a)), and $T_2$ increases with the aggregation in type II system (see Figure 16(b)). The outer sphere relaxation theory gives a theoretical explanation for these two systems. In type I system, the aggregation of MNPs cannot overcome the thermal randomization thus it's in a motional averaging (MA) model where the diffusional motion of water molecules is fast enough to average out the magnetic field generated by MNP aggregations. In this MA model (type I system, Figure 16(a)), $T_2$ is inversely proportional to the number of MNPs in aggregation and the concentration of aggregations. Thus, the more severely the MNPs aggregate, the lower the $T_2$ will be. However, once the aggregation is big enough to overcome the thermal randomization, the system becomes a static dephasing model (SD) where the free water protons contribute mainly to $T_2$. If the size of MNP aggregation is in SD regime, the $T_2$ increases with the increasing size. In this SD model (type II system, Figure 16(b)), a small number of aggregates and large space between aggregates lead to most water protons' failure to experience the magnetic field inhomogeneity [155], in this diffusion-limited case, $T_2$ increases as the size of MNP aggregates increases. This diffusion-limited case applied when larger sized MNPs are used.

Koh *et al.* [201] explored the behavior of MNP-based type I and micrometer-sized particle (MP) based type II MRSw assay systems. Both systems successfully detected the presence of Tag peptide of influenza virus hemagglutinin (IVH) and a monoclonal antibody to that peptide (anti-Tag), while type II MP based assay shows better sensitivity than type I MNP based assays. Chung *et al.* [202] reported the detection of kidney injury markers KIM-1 (kidney injury molecule-1) and Cystatin C with as low concentration of 0.1 ng/ml and 20 ng/ml respectively. MRSw also found its applications in the detection of virus such as herpes simplex virus 1 (HSV-1) [203], ions such as $Hg^{2+}$ [204, 205], $Pb^{2+}$ [206, 207] and $Cd^{2+}$ [208], and breast cancer cells, colon cancer cells, and lung cancer cells [209].

Besides its applications in disease diagnosis, MRSw has also been applied as a sensitive and rapid method for detecting foodborne pathogens such as *Listeria monocytogenes* [210], *Salmonella enterica* [211], etc. Chen et al. [212] combined the NMR and magnetic separation into a one-step platform. Based on the difference in the separation speed of small (30 nm) and large (250 nm) MNPs, both MNPs are conjugated with antibodies which specifically recognize the targets, then large MNPs are employed for separation and small MNPs are probes for MRSw sensing. Luo *et al.* [213] have demonstrated a portable MRSw-based biosensor system for the detection of *Escherichia coli* (*E. coli*) O157:H7 from drinking water and milk samples within one minute. Their device reached a detection limit of 76 cfu/mL in water samples with a linear dynamic detection range of 4 orders of magnitude.

Recently, some groups have developed miniaturized devices based on NMR for POC applications [202, 213-217], bringing down the cost, size, and sample use by orders of magnitude. As shown in Figure 16(c)&(d), a NMR system is developed by Lee *et al.* [217], it consists of a permanent magnet that generates stable magnetic field $\boldsymbol{H}_0$, a RF generator with coil close to sample to generate RF pulses, a signal receiver to amplify NMR signals and external electronics for synchronizing the different components and store the data [218].

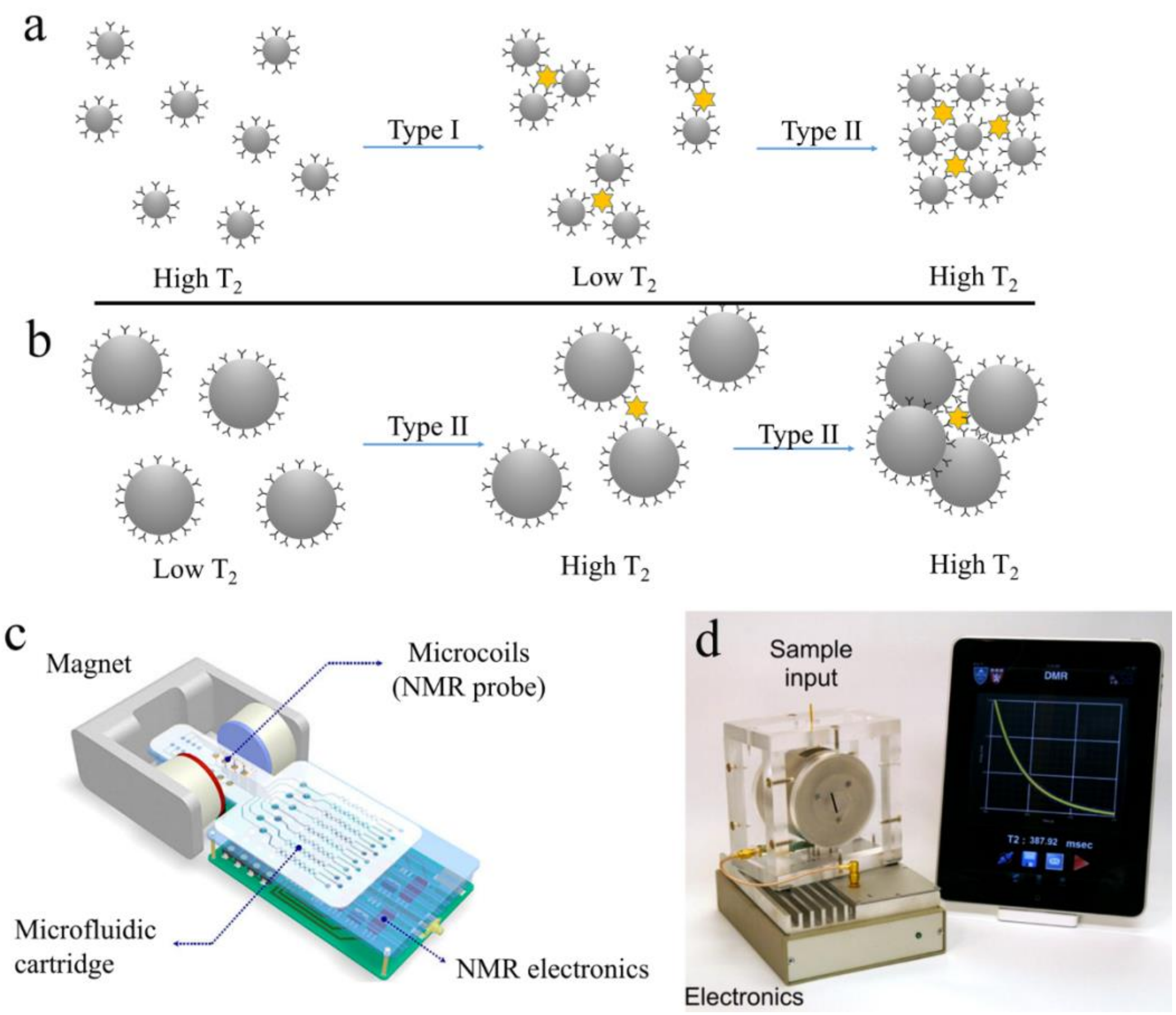

Figure 16. (a) MRSw-based detection mechanisms (type I and type II) using small MNPs. (b) MRSw-based detection mechanism (type II) using micro-sized magnetic beads. (c) Schematic diagram of the portable NMR system developed by Lee *et al.* [217]. This system consists of microfluidic networks for sample handling and mixing, an array of microcoils for NMR measurements, miniaturized NMR electronics and a permanent magnet. (d) A photograph of the portable NMR system. For user-inputs and data sharing, the system communicates with external devices. ((c) adapted with permission from reference [217], copyright 2008 Springer Nature, and (d) reprinted with permission from reference [219], copyright 2011 The Royal Society of Chemistry)

### *5.3.2 Magnetic Resonance Imaging (MRI)*

One of the most well-known applications of NMR in nanomedicine is MRI. MRI is a medical imaging technique that measures the NMR signals from protons in human bodies, and its performance can be significantly improved by administering contrast agents. The MRI contrast agents can be divided into two groups: positive ($T_1$-weighted) and negative ($T_2$-weighted) contrast agents. Positive contrast agents shorten the $T_1$ of surrounding protons and result in brighter MR images, while negative contrast agents shorten the $T_2$ of protons and lead to darker MR images. The mechanisms of MNP-based NMR have been discussed in the foregoing sections. After intravenous or oral administration, the MNPs (or USPIONs) can shorten the $T_2$ (or $T_1$) relaxation times of surrounding water protons inside various organs, leading to contrast in the MR images.

Since the first commercialization of $T_1$-weighted positive MRI contrast agent Gd-DTPA (Magnevist, Schering AG) in 1988, there have been large numbers of gadolinium-based contrast agents dominating the market (see Figure 17(a)). However, some recent studies have pointed out the concern of gadolinium-associated nephrogenic systemic fibrosis (NSF) [220-223] and FDA has issued warnings to limit the usage of gadolinium-based contrast agents [224]. Nowadays, MNPs have emerged as excellent MRI contrast agents, they have proved superior biocompatibility and safety profiles [225]. MNPs with core sizes larger than 10 nm are used as $T_2$-weighted MRI contrast agents, and ultra-small SPIONs (USPIONs) with core size less than 5 nm are reported as promising $T_1$-weighted MRI contrast agents [226-228] (see Figure 17(a)).

Based on the classical outer-sphere relaxation theory, larger MNPs with a high magnetic moment and $r_2$ relaxivities are chosen as $T_2$-weighted (negative) MRI contrast agents. There are two types of $T_2$-weighted (negative) MNP contrast agents that are clinically approved, namely: ferumoxides (Feridex in the USA, Endorem in Europe) with particle sizes of 120 to 180 nm, and ferucarbotran (Resovist) with particle sizes of about 60 nm. The principal effect of this negative contrast agents is on $T_2{}^*$ relaxation and thus MR imaging is usually performed using $T_2 - T_2{}^*$-weighted sequences in which the tissue signal loss is due to the susceptibility effects of the iron oxide core [225]. The main drawback of negative imaging agents is, however, inherently related to the contrast mechanism that they generate. MNPs produce a dark signal (a signal decreasing effect) which could be confused with other pathogenic conditions and render lower contrast compared to $T_1$ contrasted images. This main drawback has limited their clinical usages in low signal body regions, in the presence of hemorrhagic events, and in organs with intrinsic high magnetic susceptibility (such as lung) [229]. Some techniques such as spin-echo sequences [230, 231], inversion recovery ON-resonant water suppression (IRON)-MRI [232] and the usage of micron-sized iron oxide particles [233] have been proposed to overcome these challenges.

On the other hand, the $T_1$-weighted MRI contrast agents yield better imaging quality and they can effectively diagnose the normal and lesion tissues especially in blood imaging [234-236]. Kim *et al.* [237] reported one type of efficient $T_1$ contrast agent which is synthesized via the thermal decomposition of the iron-oleate complex in the presence of oleyl alcohol. These 3 nm-sized USPIONs with a high $r_1$ value of 4.78 $mM^{-1}s^{-1}$ and low $r_2/r_1$ ratio of 6.12. The synthesis methods for USPIONs include thermal decomposition, polyol, coprecipitation, or reduction precipitation and $r_1$ values vary from 2 to 50 $mM^{-1}s^{-1}$ are also reported by other groups [224, 227, 228, 238, 239].

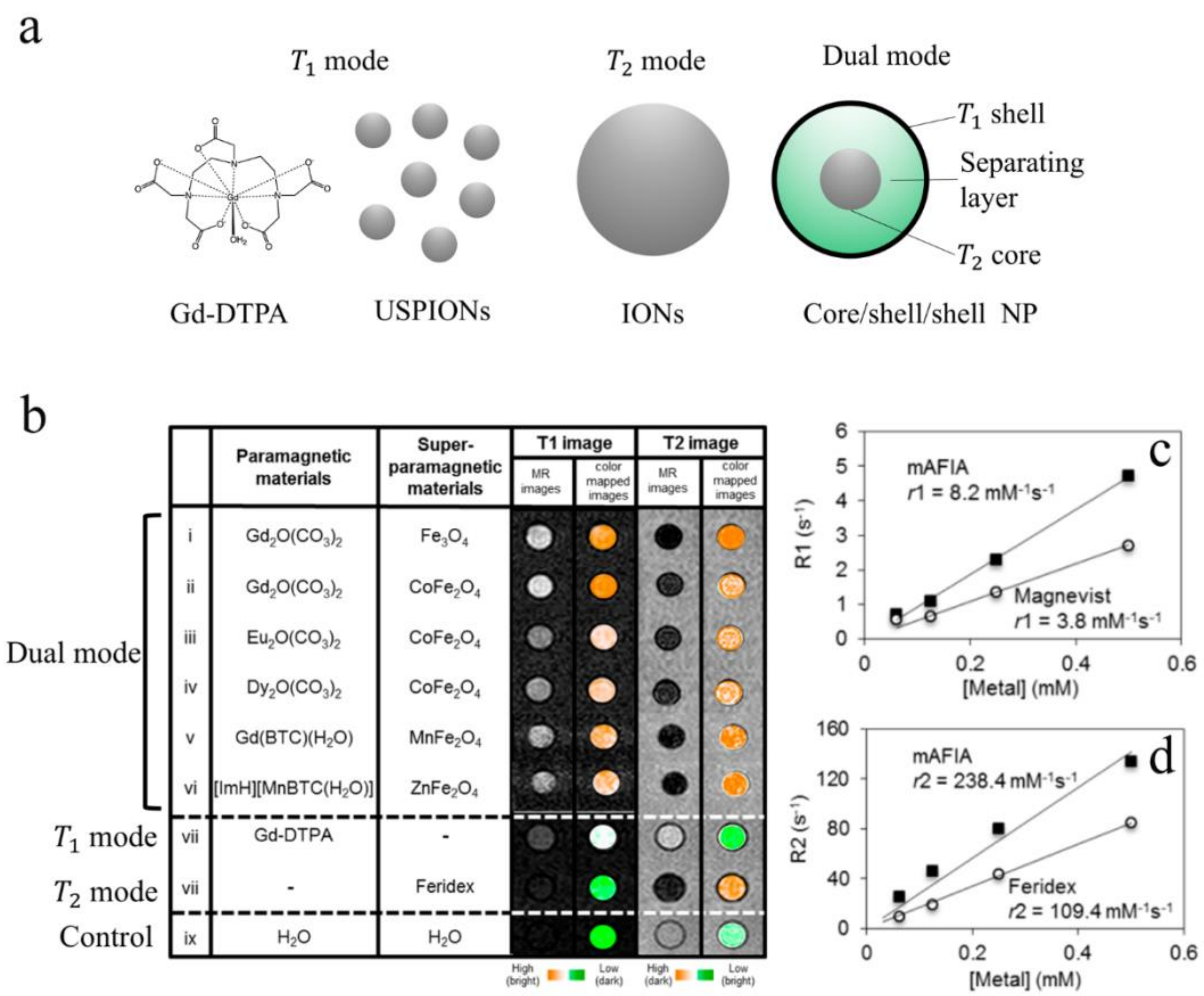

Figure 17. (a) Schematic view of MRI contrast agents: $T_1$ contrast agents such as Gd-DTPA and USPIONs, $T_2$ contrast agents such as IONs, dual mode contrast agents such as the core/shell/shell structured NPs. (b) MRI images and their color-coded images of AFIAs and conventional contrast agents. $Fe_3O_4@SiO_2@Gd_2O(CO_3)_2$ (i), $CoFe_2O_4@SiO_2@Gd_2O(CO_3)_2$ (ii), $CoFe_2O_4@SiO_2@Eu_2O(CO_3)_2$ (iii), $CoFe_2O_4@SiO_2@Dy_2O(CO_3)_2$ (iv), $MnFe_2O_4@SiO_2@Gd(BTC)(H_2O)$ (v), $ZnFe_2O_4@SiO_2@[ImH][Mn(BTC)\text{-}(H_2O)]$ (vi), Gd-DTPA (vii), Feridex (viii), and $H_2O$ (ix). Images of contrast agents are taken by using 3.0 T MRI at the identical metal concentrations. (c) & (d) Plots of $R_1$ and $R_2$ vs concentration of the metal; $r_1$ and $r_2$ of mAFIA, $ZnFe_2O_4@SiO_2@[ImH][Mn(BTC)\text{-}(H_2O)]$, are~2-fold larger than those of conventional contrast agents. ((b)-(d) reprinted with permission from reference [240], copyright 2014 American Chemical Society)

Recently, dual-mode contrast agents are emerging to eliminate the possible ambiguity of a single-mode contrast agent (either $T_1$ or $T_2$) when some of the in vivo artifacts are present, it is the combination of simultaneously strong $T_1$ and $T_2$ contrast effects in a single contrast agent (see Figure 17(b)). This dual mode contrast agent can potentially provide more accurate MRI via self-confirmation with better differentiation of normal and diseased areas. The core@shell-type $T_1 - T_2$ dual mode nanoparticle contrasts have been described by several groups [240-245]. For example, Choi *et al.* [242] reported the $MnFe_2O_4@SiO_2@Gd_2O(CO_3)_2$ core@shell@shell nanoparticles as dual-mode MRI contrast agents, where the $SiO_2$ layer was used to modulate the magnetic coupling between $T_1$ contrast material $Gd_2O(CO_3)_2$ and $T_2$ contrast material $MnFe_2O_4$. By increasing the $SiO_2$ layer thickness (varies from 4 nm to 20 nm), the $r_1$ value increased from 2.0 to 33.1 $mM^{-1}s^{-1}$, and the $r_2$ value decreased from 332 to 213 $mM^{-1}s^{-1}$. In Figure 17(c)-(d), Shin *et al.* [240] measured different combinations of core@shell@shell dual mode artifact filtering nanoparticle imaging agent (AFIA) [$T_2$ core (superparamagnetic nanoparticle) @$SiO_2$@$T_1$ shell (paramagnetic material)] and demonstrated its superior relaxivities ($r_1$ and $r_2$) than Magnevist ($T_1$ contrast agent) and Feridex ($T_2$ contrast agent). Other structures such as iron core with ferrite shell nanoparticles [246, 247], ultrasmall mixed gadolinium-dysprosium oxide (GDO) nanoparticles [248] and core@shell structured manganese-loaded dual-mesoporous silica spheres (Mn-DMSSs) [249] have also been reported as $T_1 - T_2$ dual mode contrast agents.

## 5.4 Superparamagnetism-based Diagnostics

### *5.4.1 Brownian Relaxation-based Assay*

As is aforementioned in section 2.3, superparamagnetic nanoparticles exhibit non-linear magnetic response curve with zero coercivity. Under an external AC magnetic field $H$, the magnetic moment inside superparamagnetic nanoparticle tends to align with the field while this process is countered by Néel and Brownian relaxations that randomize its magnetic moment [250, 251]. For superparamagnetic iron oxide nanoparticles (SPIONs) dispersed in liquid solution, the Brownian relaxation is the dominating obstructing factor when the magnetic core size is above 20 nm while Néel relaxation dominates when core size is below 20 nm [79, 252]. For those Brownian relaxation-dominated SPIONs, the extent of the disorder is linked directly to environmental conditions [253], such as temperature [254, 255], viscosity [79, 256-258], and the ligand binding states [80, 259-262]. In this section, we focus on the Brownian relaxation dominated SPIONs for immunoassay applications. For Brownian relaxation dominated SPIONs, the effective relaxation time is expressed as $\tau \approx \tau_B = \frac{3\eta V_h}{k_B T}$. SPIONs are functionalized with ligands (such as aptamers, proteins, antibodies, etc.) which can specifically recognize and bind to target analytes from minimally processed biofluid samples, this successful binding process increases the hydrodynamic size of SPIONs thus the effective relaxation time. Hence, a larger phase delay between its magnetic moment and the AC magnetic field is detected, and the magnitudes of harmonics are attenuated as a result (see Figure 18(h)&(i)).

In 2006, Krause group [263] and Nikitin group [264] have independently proposed a mixing frequency method based Brownian relaxation dominated SPIONs for immunoassay applications. As shown in Figure 18(a)-(g), two drive fields $H = A_H cos(2\pi f_H t + \varphi_H) + A_L cos(2\pi f_L t + \varphi_L)$ are applied to drive SPION suspension, the response signal generated at $f_H \pm 2f_L$ (3rd harmonics), $f_H \pm 4f_L$ (5th harmonics), etc., are collected. These harmonics are highly specific to the nonlinearity of the magnetization curve of the SPIONs. The amplitude of low frequency AC field, $A_L$, is chosen to periodically switch on and off the capability of SPIONs to further magnetization, while the high frequency AC field is used to modulate the harmonics into high frequency regions where the pink noise ($1/f$ noise) is lower. Due to the time-varying magnetization of the SPIONs, a time-variant voltage is induced in the receive coil due to the SPION harmonic response (Figure 18(b)). On the other hand, Rauwerdink and Weaver et al. [255, 260, 262, 265] proposed a mono-frequency method based Brownian relaxation dominated SPIONs for immunoassay applications, its setup is similar to the 1D MPI: they applied an AC magnetic field $H = Acos(2\pi f t + \varphi)$ to drive SPIONs, the nonlinear magnetic response induces higher harmonics at frequencies of $3f$ ($3^{rd}$ harmonic), $5f$ ($5^{th}$ harmonic), etc., the information such as SPION-target analyte binding states can be extracted from these harmonics in the same way as we described before.

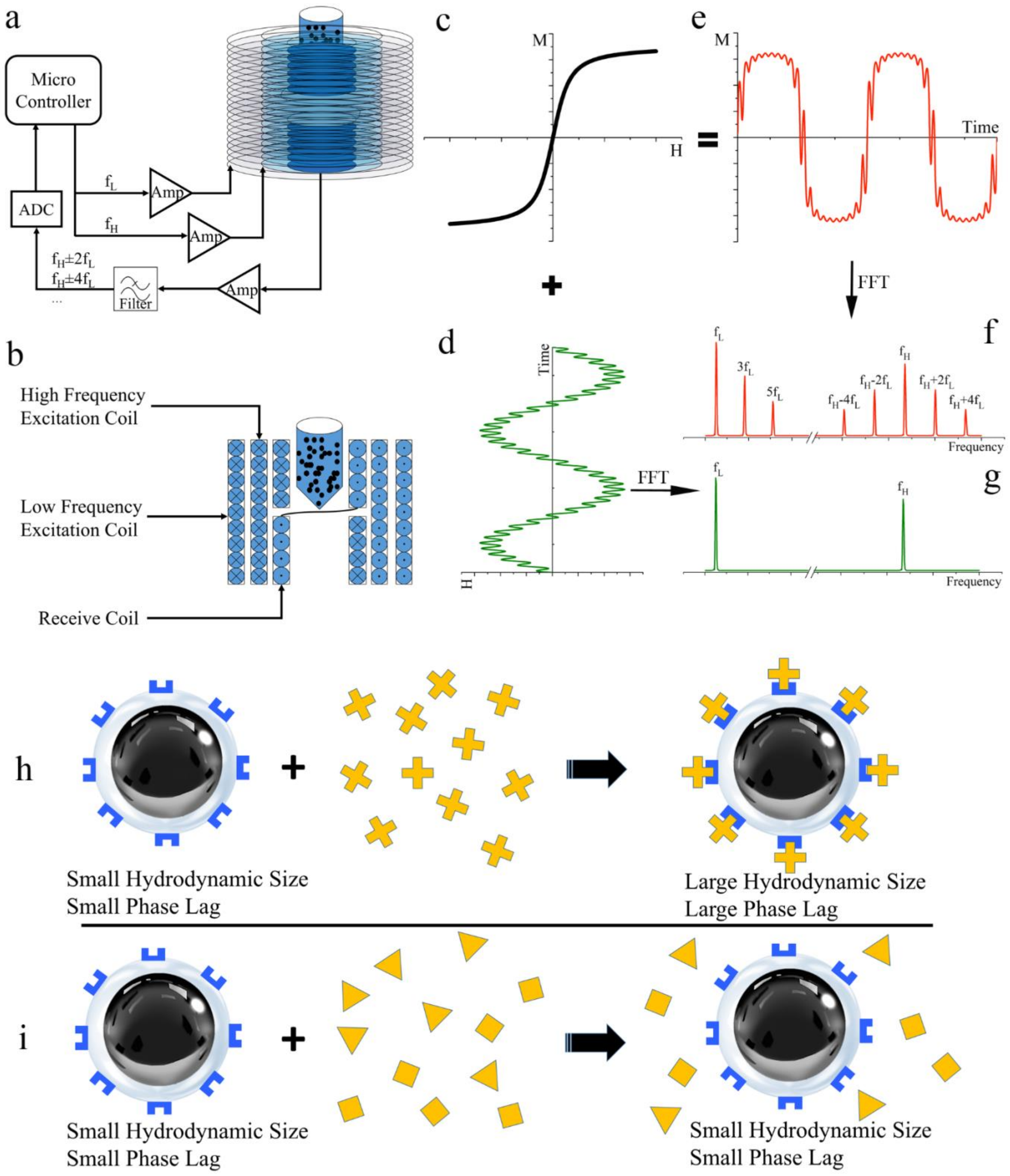

Figure 18. (a) Schematic setup of mixing frequency method-based magnetometer. Microcontroller with electronics part provide sine waves to drive the excitation coils, the generated signals are picked up by receive coil and harmonics are filtered and read into the microcontroller via the ADC. (b) Sectional view of the coils. (c)-(g) SPIONs are exposed to two AC magnetic field with frequencies of $f_H$ and $f_L$ (d). The drive field spectrum in (g) exhibits two distinct lines. Due to the nonlinear magnetization curve of SPIONs (c), the resulting time-

dependent magnetization (e) saturates at higher fields, leading to higher harmonics and frequency mixing components in the Fourier-transformed response signal (f). (h) SPIONs are functionalized with ligands that will specifically bind to target analytes, this binding process increases the SPION's hydrodynamic size as well as phase lag. (i) The ligands on SPIONs blocks them from non-specific binding and prevents false signal change.

Since the amplitudes of harmonics increase with the number of SPIONs within the fluid sample, so phase lags [80, 266] and harmonic ratio [265] (magnitude ratio of 5$^{th}$ over 3$^{rd}$ harmonic) are used as concentration-independent metrics to characterize the relaxation time as well as the binding state of SPIONs.

Orlov *et al.* [267] applied the method of registration of SPIONs by their nonlinear magnetization to a novel immunoassay on 3D fiber solid phase for detection of staphylococcal toxins in unprocessed samples of virtually any volume (see Figure 19(a)). This 3D fiber structure serves as a reaction surface to selectively filter antigens as well as accelerate reagent mixing. Two different setups are reported, the limits of detection (LOD) reaches to 4 and 10 pg/mL for toxic shock syndrome toxin (TSST) and staphylococcal enterotoxin A (SEA), respectively, by using 30 mL samples, in 2 hours. Tu *et al.* [80] reported a real-time monitoring of the increase of phase delay in 3$^{rd}$ harmonic due to the binding of goat anti-human IgG to protein G coated SPIONs (see Figure 19(b)). Which is the first experimental demonstration of real-time detection of analyte binding process. Zhang *et al.* [262] demonstrated a rapid measurement (within 10 s) using SPIONs that conjugated with anti-thrombin ssDNA aptamers for the detection of thrombin, a LOD of 4 nM is reached (see Figure 19(c)).

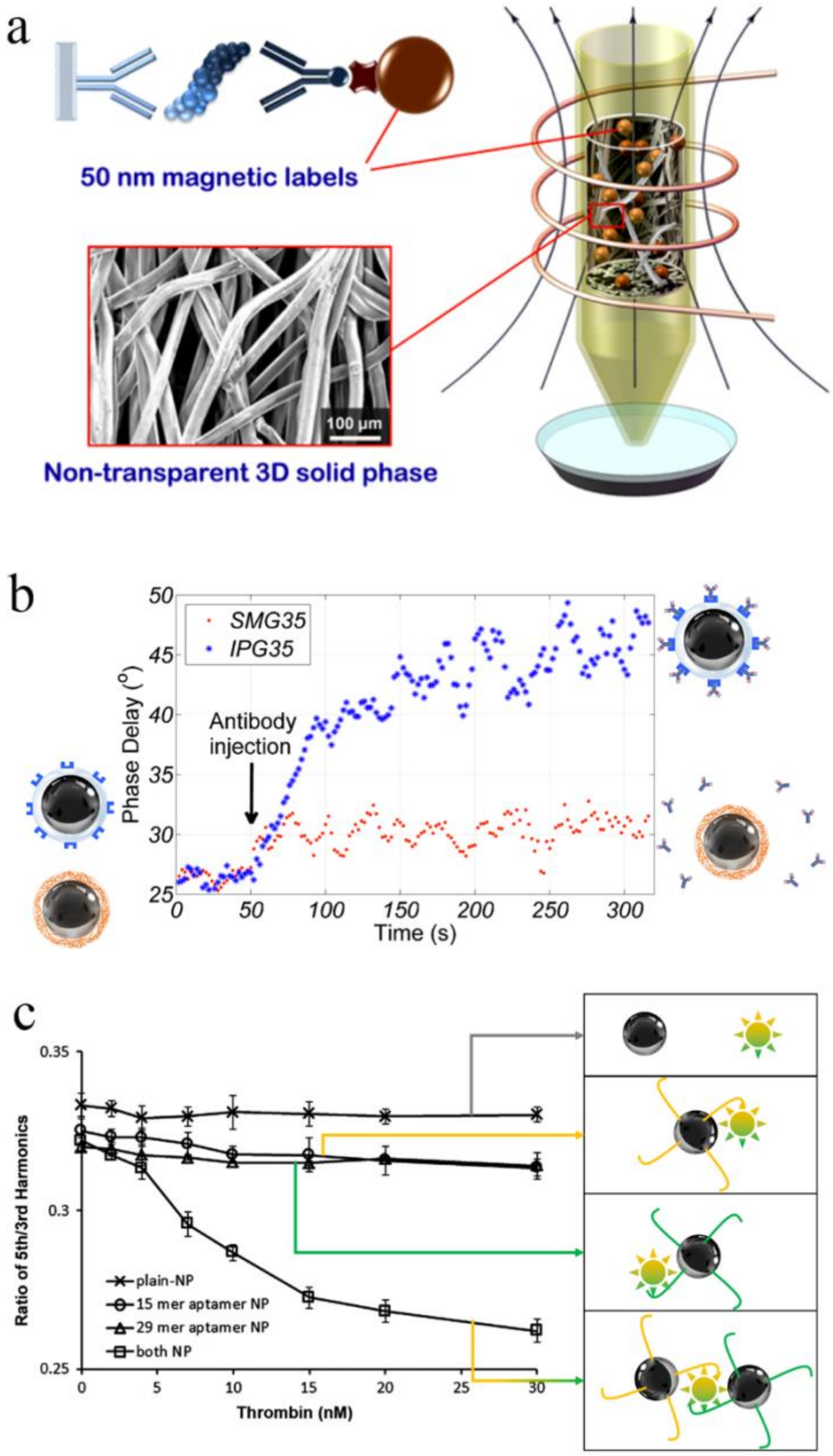

Figure 19. (a) Top left: Magnetic immune-sandwich on 3D fiber filters used as a solid phase. Bottom left: SEM surface morphology of the cylindrical 3D fiber filters. Right: Detection of SPIONs by their nonlinear response at combinatorial frequencies from the whole volume of the 3D solid phase located inside a pipet tip. (b) Real time measurement of phase delay of IPG35 in blue curve and control sample SMG35 in red curve, with antibody injection at the 50th second. IPG35 are SPIONs coated with protein G that will specifically bind to IgG antibodies, this binding process gradually increase the hydrodynamic size of IPG35 SPIONs. SMG35 are SPIONs coated with PEG layer which prevents the binding of IgG onto the SPION surface, as a result, hydrodynamic size of SMG35 barely changed. IPG35 and SMG35 are SPIONs with average hydrodynamic sizes of 35 nm, thus, they exhibit same initial phase delays under the same driving fields. (c) Ratio of $5^{th}$ to $3^{rd}$ harmonic of 2 populations of SPIONs, conjugated with 15 mer and 29 mer anti-thrombin aptamer, harmonic ratios are measured as a function

of thrombin concentration. Three control samples are applied in this experiment to verify the specificity of the detection: SPIONs without aptamer functionalized, and only one population of SPION functionalized with either 15 mer or 29 mer anti-thrombin aptamer with the increasing concentration of thrombin added. ((a) adapted with permission from reference [267], copyright 2012 American Chemical Society, (b) adapted with permission from reference [80], copyright 2011 AIP Publishing LLC, (c) adapted with permission from reference [262], copyright 2013 Elsevier)

### 5.4.2 Magnetic Particle Imaging (MPI)

Magnetic particle imaging (MPI) has shown great promise in many clinical applications ranging from angiography to cancer theranostics and molecular imaging. This technology was first reported in 2005 by Gleich and Weizenecker [268], which exploits the nonlinearity of magnetization curves of SPIONs and the fact that their magnetization saturates at some magnetic field strength. It's also a safer imaging method especially for chronic kidney disease (CKD) patients for whom iodine contrast agents are toxic. The dynamic 2D MPI and the first prototype of 3D real-time MPI were proposed by Gleich and Weizenecker *et al.* [269, 270], pushing a huge step toward its medical applications. A typical modality of 3D MPI system is shown in Figure 20, an AC modulation magnetic field $\boldsymbol{H}(\boldsymbol{t})$ (it's also called drive field) excites SPIONs and the nonlinear magnetic response $\boldsymbol{M}(\boldsymbol{t})$ is induced and picked up by a receive coil. This magnetic response signal $\boldsymbol{M}(\boldsymbol{t})$ contains not only the modulation field frequency $f$, but also a series of harmonic frequencies $3f, 5f, 7f$ etc. Higher harmonics instead of the fundamental frequency are used for analysis to avoid picking up the applied field. These higher harmonics can be easily separated from the received signal by filtering. In addition to the modulation field, an additional gradient field (also called selection field) is applied to localize received signals. This selection field features a field free point (FFP) where its magnitude is zero and increases towards the edges (see Figure 20(b)), this FFP is scanned through the sample following the scan trajectory. SPIONs in the FFP (red region in (b) and red curves in (c)) produce signals containing higher harmonics while those in the saturation region (the grey region in (b) and grey curves in (c)) contribute negligible signals. Tomographic MPI images are generated by scanning the FFP through the volume of interest.

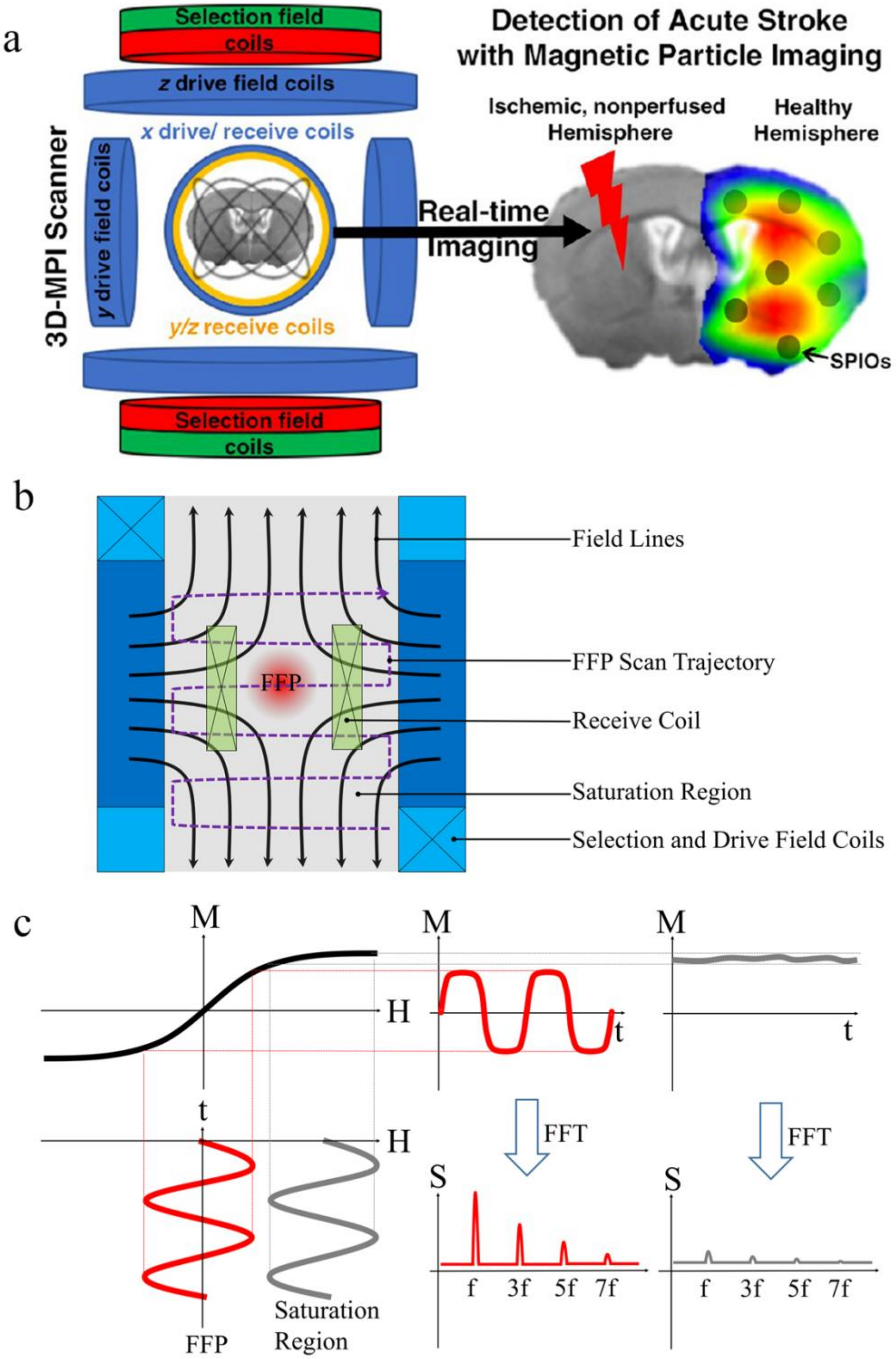

Figure 20. (a) Schematic view of 3D MPI scanner. The biological sample was inserted into the x drive/receive coil cylinder. The selection field is generated by both the coil pair in the z-direction. The drive field coils can move the FFP in all three spatial directions. Only the SPIONs in the instantaneous location of the FFP generate MPI signals. For signal reception, each spatial component of the magnetization is detected by a respective receive coil. In the x-direction, the drive field coil is also used for signal reception. (b) Illustration of FFP scanning and distribution of magnetic field. SPIONs locate in the saturation region contribute negligible MPI signals. (c) The response of the SPIONs to external magnetic fields. When the modulation field $\boldsymbol{H}$ with a frequency of $f$ is applied, the magnetic response $M(t)$ exhibits higher harmonics ($3f, 5f, 7f, etc.,$) as shown in the Fourier Transformed signal (red curves). These higher harmonics are used for MPI image generation. When an additional selection field is added (grey curves), the SPIONs are magnetically saturated and the magnetic response $M(t)$ does not significantly change. ((a) reprinted with permission from reference [271], copyright 2017 American Chemical Society)

MPI is an ideal tool for angiography, cellular and targeted imaging, cancer imaging, as well as imaging major organs and the finer coronary arteries, it allows 3D visualization, real-time imaging, and moreover, it's radiation-free [269, 272]. MNP tracers can only be detected indirectly in MRI due to their effect on the magnetic resonance signal of the nuclear spin magnetization. On the other hand, MPI directly detects MNP (SPION) tracers based on the nonlinear magnetization response. Thus, MPI provides the distribution of tracer SPIONs deep in body with relatively high temporal and good spatial resolution, and the quality of MPI images are largely dependent on the properties of SPION tracers (core diameter, shell thickness) and scan speed [28, 273-275].

In recent years, the development of customized SPIONs for optimized MPI is pursued with great effort. Dhavalikar *et al.* [274] reported that the MPI signal strength is related to the cubic power of the SPION core diameter $D$ as long as superparamagnetic characteristics are retained. In addition, the MPI spatial resolution is heavily dependent on the SPION's magnetic response curve and the gradient selection field. To be specific, a steeper magnetization curve confines a smaller FFP, similarly, a larger gradient selection field shrinks FFP, so a better spatial resolution is achieved [270, 276]. Since the SPION's magnetic response to a time-variant magnetic field is governed by Brownian or/and Néel relaxations, the spatial resolution is highly dependent on the SPION's relaxation characteristics.

In the first generation of MPI systems, the drive fields were around 10-20 mT in magnitude and frequency of 25 kHz. Theoretically, an ideal SPION core size of 30 nm was proposed for this drive field setting [268]. However, Lüdtke-Buzug et al. [277] evaluated the commercially available SPION contrast agents for MRI regarding their MPI performances, and he Resovist (Bayer Pharma AG), with core size much smaller than 30 nm, showed the best MPI performance over other tracers. Later research found that the Resovist SPIONs form aggregates and each aggregate still behaves like one SPION with a core size equivalent to 24 nm [29]. Besides this SPION aggregation factor, other factors such as the SPION size distribution, anisotropy constant, morphology, shell thickness, inter-particle interactions also affect the MPI performance [278]. To sum up, the properties of SPION, as well as the interaction with the environment must be considered and modified for different drive field settings [33, 279, 280].

## 5.5 Surface Enhanced Raman Spectroscopy (SERS) Systems

Raman scattering results from the radiation emitted by molecules or atoms after the bombardment of a primary radiation [281]. The absorption of energy due to inelastic scattering of photons causes rotational and vibrational changes, which in turn leads to the change of the molecular dipole-electric polarizability. Raman Spectroscopy (RS) often uses a non-ionizing laser as the excitation source, and each peak in its signal can provide information on a specific chemical bond. While RS is a non-invasive and label-free technique, its signal is rather weak with a large fluorescence background. Moreover, there is also a limitation on the penetration depth [282].

In pursuit of stronger signal magnitude, SERS was firstly introduced by Fleishmann *et al.* [283]. The substrates of the SERS are nano-noble metals. As the target molecules are absorbed to the surface of the substrate, there will be a significant increase in the Raman signal due to the excitation of the surface plasmons, which makes it possible for SERS to detect trace biomarkers even at a large distance from the target tissues [281]. However, when only plasmonic NPs (eg., Au or Ag NPs) are used, the non-uniform attachment of NPs on bacteria makes it difficult to quantitatively measure the concentration. Detection of targets with very low concentrations is also impossible due to the lack of approaches to concentrate the collected samples. To solve this problem, Zhang *et al.* [284] developed multifunctional magnetic-plasmonic $Fe_2O_3$@Au core@shell NPs. With the magnetic components inside the NPs, the plasmonic properties can be tuned by changing the inter-particle distance via the external magnetic field (see Figure 21) [285]. Furthermore, MNPs can also facilitate the fast concentration of bacteria cells under a point magnetic field. It was demonstrated that the plasmonic MNPs can concentrate the bacteria to a level of ~ 60 times higher than non-concentrated samples, which greatly enhanced the SEPR signal.

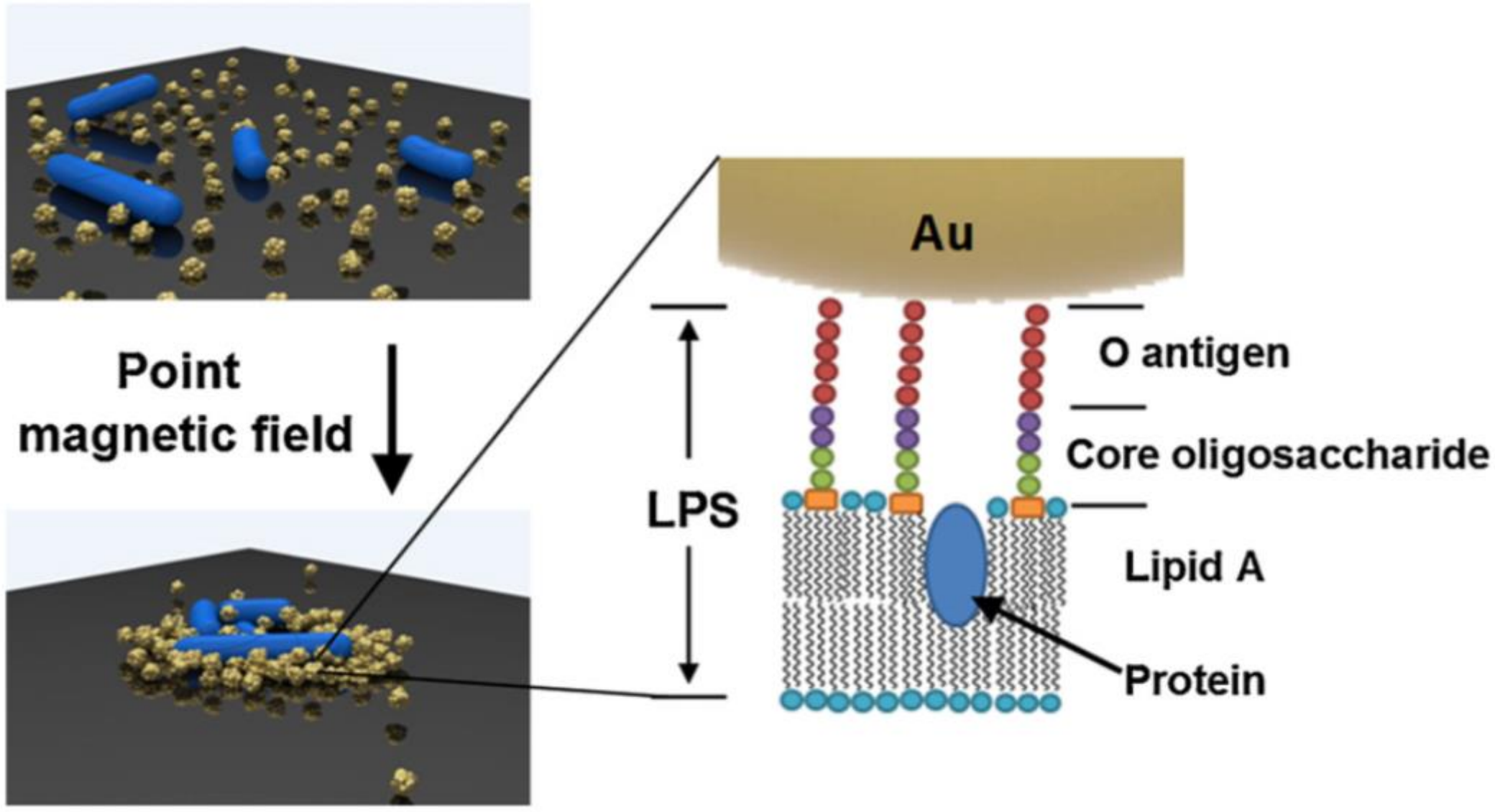

Figure 21. Schematics of the condensation process of Au MNPs and bacteria(left) and the biomolecular characteristics of the bacterial cell wall that can possibly be detected by SERS (right). (Reprinted with permission from reference [284], copyright 2011 Elsevier)

Due to their superior properties in SEPR detection, plasmonic MNPs were employed in the detection of a wide variety of pathogen bacteria [286-288] and also in immunomagnetic separations [289]. For example, Guven *et al.* [286] used ion-oxide MNPs with Au nanorods (NRs) as substrates to detect *E. coli*, achieving a limit of detection at 8 cfu/mL. Drake *et al.* [287] developed the same MNPs together with Au NPs for the detection of *S. aureus*. The detectable concentration was shown to be as low as 1 cell/mL.

For the convenience of the readers, different MNP-based immunoassay techniques that have been reviewed in this paper are summarized in Table 1.

Table 1. Comparison of different MNP-based immunoassay techniques

| Sensor Type | MNP | Assay Time | Target Analyte | Limit of Detection (LOD) | Evaluated Matrices | Ref. |
|---|---|---|---|---|---|---|
| **GMR** | 50 nm MNP | 30 min | Pregnancy-associated plasma protein A (PAPP-A) | 1 ng/mL | Blood serum | 168 |
| | | | Proprotein convertase subtilisin/kexin type 9 (PCSK9) | 433.4 pg/mL | | |
| | | | Suppression of tumorigenicity 2 (ST2) | 40 pg/mL | | |
| **GMR** | 50 nm MNP | 10 min | Influenza A Virus (IAV) Nucleoprotein (NP) | 15 ng/mL | Phosphate buffered saline (PBS) | 171 |
| | | | Purified H3N3 Virus | 125 $TCID_{50}$/mL | | |
| **GMR** | 50 nm MNP | 1 hr | DNA with NRAS and BRAF mutations | NA | 20×saline sodium citrate (SSC) | 173 |
| **MTJ** | 250 nm MNP | 30 min | DNA from Listeria, HEV and Salmonella | 1 nM | Phosphate buffer | 187 |
| **MTJ** | 12 nm MNP | 12 hr | Single strand DNA | 10 nM | PBS | 290 |
| **µHall** | 20 nm MNP | 40 min | *S. aureus* | 10 bacteria | Staphylococcus broth | 194 |
| **µHall** | 250 nm MNP | 24 hr | Oligonucleotides | NA | NaCl+HCl+EDTA | 291 |
| **MRSw** | 30 nm MNP | NA | Kidney injury molecule-1 (KIM-1) | 0.1 ng/mL | Urine | 202 |
| | | | Cystatin C | 20 ng/mL | Urine | |
| **MRSw** | 50 nm MNP | 30 min | *S. enterica* | $10^3$ cfu/mL | Milk | 211 |
| **MRSw** | 30 nm MNP | 30 min | *S. enterica* | $10^4$ cfu/mL | Purified | 212 |
| | | | *S. enterica* | $10^3$ copy/mL | Purified | |
| **MRSw** | 30 nm MNP | 90 min | Newcastle disease virus | 5 fM | Urine | 292 |
| **MS-MRSw** [a)] | 30 nm MNP & 250 nm MB | 30 min | miR-21 from tumor cells | $10^2$ cfu/mL | Purified | 212 |
| | | | *S. enterica* | $10^2$ copy/mL | Purified | |
| **Brownian Relaxation-based Assay** | 50 nm MNP | 10 sec | Newcastle disease virus | 150 pM or 0.075 pmole | Purified | 262 |
| | | | Streptavidin | 4 nM or 2 pmole | Purified | |
| | | | Thrombin | 100 pM or 0.05 pmole | Purified | |
| **Brownian Relaxation-based Assay** | 50 nm MNP | 2 h | Anti-thrombin aptamers | 4 pg/mL, 30 mL | Milk | 267 |
| | | | Staphylococcal enterotoxin A (SEA) | 10 pg/mL, 30 mL | Milk | |
| | | 25 min | Toxic shock syndrome toxin (TSST) | 0.1 ng/mL, 150 µL | Milk | |
| | | | SEA | 0.3 ng/mL, 150 µL | Milk | |
| **SERS** | 200 nm Au-MNP | 3 hours | 4-mercatopyridine | Lower than 1 nM | DI water | 284 |
| **SERS** | 200 nm Au-MNP | NA | *E. coli*<br>*P. aeruginosa* | $2*10^5$ cfu/mL | DI water | 284 |
| **SERS** | 100 nm Silica-coated MNP | 2 hours | *S. aureus* | $10^3$ cfu/mL | Spinach wash | 288 |
| **MS-SERS** [b)] | 500 nm MB | 2 hours | TSST | 1 pg/mL | Purified | 293 |

[a).] Magnetic Separation and Magnetic Relaxation Switching assay

[b).] Magnetic Separation and Surface-enhanced Raman Scattering assay

# 6 MNPs for Therapeutic Applications

## 6.1 Drug Delivery

The concept of using MNPs/magnetic microparticles as carriers for targeted therapeutic applications was first proposed in the late 1970s by Senyei and Widder *et al.* [294, 295]. The therapeutic agents (such as cytotoxic drugs for chemotherapy and DNA for genetic therapy) are attached to the surface of the magnetic carrier or encapsulated within the polymer matrix which contains the magnetic carrier. These particle-drug complexes are injected into the subject either via oral or intravenous, then an externally applied high-gradient magnetic field guides and concentrates these complexes to the target sites (see Figure 22(a)). Once the complexes are concentrated at the target organ in the body, the therapeutic agents are released via enzymatic cleavage of the cross-linking molecules, charge interactions, pH change, degradation of the polymer matrix, or heating up the particles [296-299]. This MNP-based targeting method effectively reduces the side-effects brought by the non-specific nature of chemotherapy, which attacks healthy cells in addition to primary targets. Furthermore, by accurately administering and delivering therapeutic agents to target sites, it reduces the systemic distribution of cytotoxic compounds and enables effective treatment with a lower dosage.

SPION is one of the most commonly used magnetic carriers for drug delivery, it's often coated with an organic matrix such as polysaccharides, fatty acids, or other polymers to improve the colloidal stability [199, 300-302]. SPIONs exhibit zero remnant magnetization upon the removal of the external field, which avoids the aggregations and facilitates their excretion from the body. Magnetic targeting is based on the theory that, a translational force will be exerted on the particle-drug complex in the presence of a magnetic field gradient, thus trapping the complex at the target site and pulling the complex towards the magnet [296, 303]. A successful trapping of the particle-drug complex at the target site is largely dependent on the blood flow rate, surface characteristics of magnetic carriers, magnetic properties of particles, the magnetic field strength as well as magnetic field gradient.

Since the magnetic field strength decays rapidly with distance to the inner sites of the body become more difficult to target, which is one of the main barriers that impeding the scale-up of magnetic targeting from small animals to humans [297]. Some groups have proposed to implant magnets (i.e. a magnetizable intraluminal stent, seed, etc.) in the body near the target site to circumvent this problem [304-309]. Nacev *et al.* [310] reported pulsed magnetic fields to exploit the rotational dynamics of ferromagnetic rods and attain well-like curvature of the magnetic potential energy (see Figure 22(b)). Using this method, they focused a disperse concentration of ferromagnetic rods to a central target between eight external electromagnets (see Figure 22(c)&(d)).

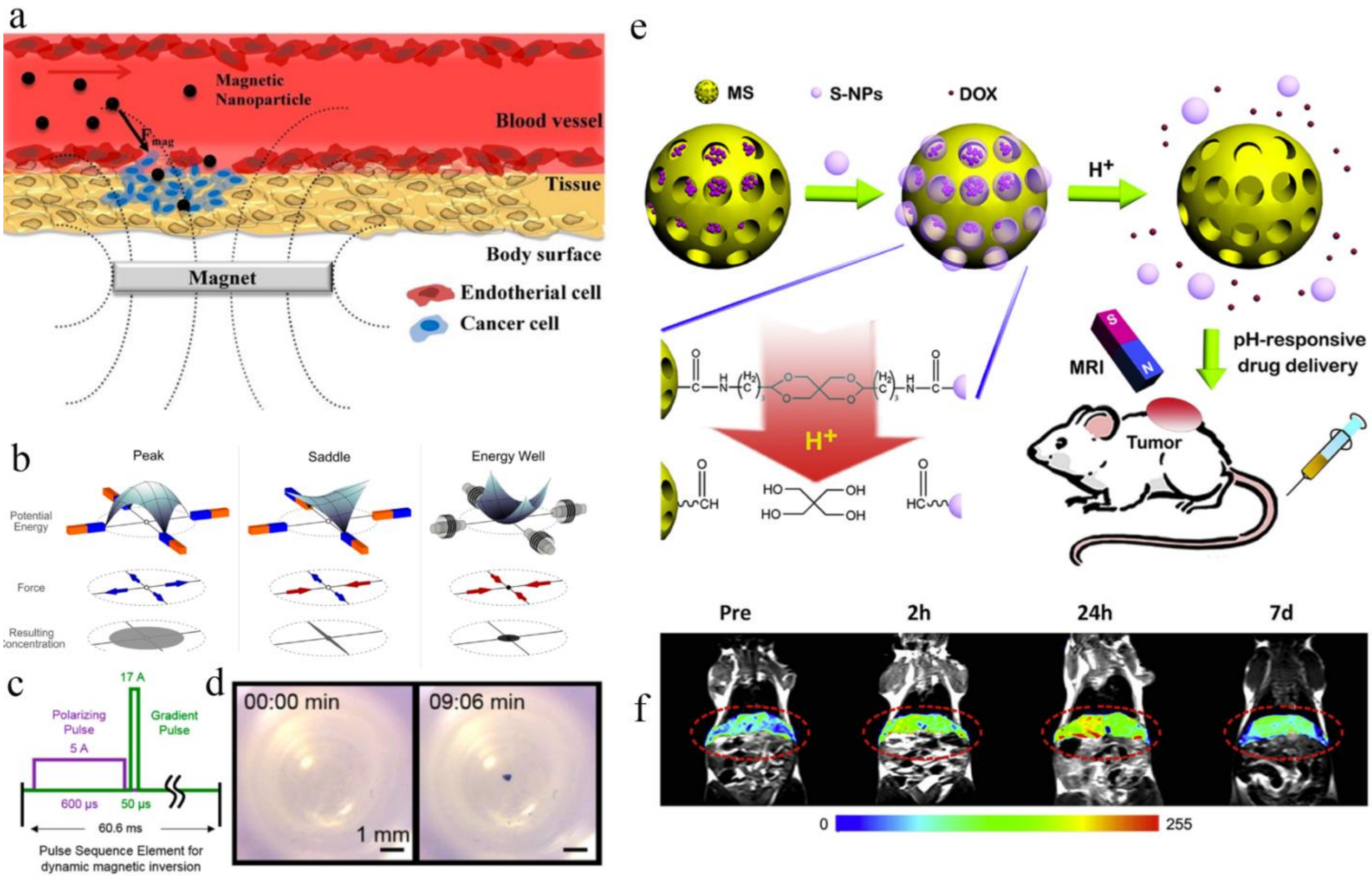

Figure 22. (a) Schematic representation of magnetic nanoparticle-based drug delivery system. (b) How forces generated from a magnet configuration affect particle concentration. A magnet configuration creates a magnetic potential energy surface (top row) that generates the magnetic forces. Magnetic forces (middle row) shape particle concentrations (bottom row). Particles will move from locations of high-energy states (white) to low-energy locations (black). Equivalently, particles will move due to either divergent forces (blue arrows) or convergent forces (red arrows). By applying Earnshaw's theorem to static magnetic fields, only unstable static magnetic potential energy configurations were theorized to be possible, e.g., (left) a peak energy configuration and (middle) a saddle; (right) through the use of pulsed magnetic fields, a magnetic potential energy well is generated which is capable of concentrating particles to a central target. (c) Pulse sequence element for inverting the energy surface

of ferromagnetic rods and concentrating them at the center of the sample area. This pulse sequence element is repeated many times for the four directions to push and concentrate the rods to the center. (d) Focusing of ferromagnetic rods to a central target. Four snapshots of concentrating cobalt rods to the center of the sample area using dynamic magnetic inversion. The rods began optically undetectable and dispersed throughout the region. After 09:06 min, the rods were concentrated at the center of the sample area. Video available as Supporting Information. (e) Schematic illustration of the MEMSN system for pH-responsive drug delivery and magnetic resonance imaging. (f) In vivo MR imaging of the mice after intravenous injection of MS@S-NPs at a different time period, the red circles point the liver. ((a) reprinted with permission from reference [311], copyright 2009 Elsevier, (b)-(d) reprinted with permission from reference [310], copyright 2014 American Chemical Society, (e)&(f) reprinted with permission from reference [298], copyright 2015 Elsevier)

Besides the magnetic field issue, there are several other obstacles impeding the delivery of drug to target sites, such as unspecific uptake by healthy cells, rapid clearance, and aggregation during the circulation. To overcome these barriers, an ideal drug carrier should have the following properties: a protection layer to keep the carrier stable and avoid nonspecific cell uptake, tumor-targeting ability to accumulate at tumor sites, ligand-mediated cell adhesion for efficient cellular entry, and effective drug encapsulation to permit the drug release inside tumor cells only [312].

In recent years, magnetic carriers with multiple functions including targeted anticancer drug delivery, hyperthermia, and imaging have been an ongoing hot topic [313-319]. Hervault *et al.* [317] reported a dual-pH and thermo-responsive magnetic carrier with combined magnetic hyperthermia and anticancer drug delivery. In their study, a pH- and thermo-responsive polymer shell is coated onto SPION, then the anticancer drug doxorubicin hydrochloride (DOX) is conjugated via acid-cleavable imine linker. This complex provides advanced features for the targeted delivery of DOX via the combination of magnetic targeting, and dual pH- and thermo-responsive behavior which offers spatial and temporal control over the release of DOX. Chen et al. [298] reported a pH-responsive nano-gated multifunctional envelope-type mesoporous silica nanoparticle (MEMSN) which is capable of magnetic drug delivery and *in vivo* MRI. They immobilized acetals on the surface of mesoporous silica and coupled to ultra-small lanthanide doped nanoparticles ($NaGdF_4$) coated with $TaO_x$ layer (S-NPs) as gatekeeper (see Figure 22(e)). The anticancer drug DOX is locked in the pores and its burst release can be achieved under acidic environment on account of the hydrolyzation reactions of acetals. This MEMSN system has demonstrated *in vivo* MRI (see Figure 22(f)), passive tumor targeting, increased tumor accumulation, and it can be harmlessly metabolized and degraded into apparently nontoxic products within a few days. Ye *et al.* [316] reported a nanocarrier system with multiple imaging agents and an anticancer drug, they encapsulated inorganic imaging agents of SPION, manganese-doped zinc sulfide (Mn:ZnS) quantum dots (QDs) and the anticancer drug into poly(lactic-co-glycolic acid) (PLGA) vesicles via an emulsion-evaporation method. Their PLGA vesicles exhibit high ${r_2}^*$ ($523\ s^{-1} mM^{-1}\ Fe$) relaxivity and greatly enhanced ${T_2}^*$-weighted MR imaging contrast. The Mn:ZnS QDs are used for fluorescence imaging to investigate the interaction between PLGA vesicles and cells.

## 6.2 Hyperthermia Therapy

Magnetic hyperthermia is a promising method for the treatment of cancer, currently in clinical trials [320-323]. In hyperthermia therapy, an AC magnetic field is remotely controlling MNPs to induce local heat, provoking a local temperature increase in the target tissues where the tumor cells are present. The specificity of this technique is based on that tumor cells are vulnerable to temperatures above 42 °C, at which the natural enzymatic processes that keep tumor cells alive are destroyed, so allowing the selective killing [324]. The AC field used in magnetic hyperthermia is in the range of radio-frequency, between several kHz and 1 MHz, which is completely healthy and shows enough penetration depth to access inner organs/tissues in the body. Several heating mechanisms are possible in magnetic hyperthermia, namely, hysteresis loss, Brownian and Néel relaxations (susceptibility loss), and viscous heating. In comparison to these aforementioned magnetic losses, the eddy current is only significant in materials at centimeter scale and it's negligibly small in MNP hyperthermia applications.

For multi-domain MNPs, the hysteresis loss is determined by integrating the area of hysteresis loops, which is a measure of energy dissipated per cycle of magnetization reversal. This type of heating attributes to the shifting of domain walls and it's strongly dependent on the AC field amplitude and its magnetic prehistory. As the MNP size reduces to the single domain, hysteresis heating is accomplished by rotating the magnetic moment of each MNP against the energy barrier. As the MNP size further reduces to superparamagnetic NP where the thermal fluctuation becomes comparable to the energy barrier, the relaxation mechanism plays a major role in magnetic hyperthermia. During Brownian relaxation process the thermal energy is delivered through shear stress in the surrounding fluid (viscous heating) and during Néel relaxation process the thermal energy is dissipated by the rearrangement of atomic dipole moments within the crystal [321]. For those superparamagnetic NPs in a liquid environment, each particle maintains a constant magnetic moment and its orientation is determined by the effective anisotropy of the particle. An external magnetic field switches the moment from its preferred orientation and the relaxation of the moment back to equilibrium state releases thermal energy which results in local heating [325]. Susceptibility loss is associated with Brownian and Néel relaxations, recall equation (11), the specific absorption rate (SAR), $P\ (W \cdot m^3)$, is related to $\chi''$ according to $P = \frac{1}{2}\mu_0 \omega \chi'' {H_0}^2$, the global maximum $P$ is reached when $\omega\tau = 1$ (SAR is related to the specific loss power, namely, $SLP\ (W \cdot g^{-1})$, by the mean mass density of particles). The strong size dependence of relaxation time results in a very narrow maximum of the loss of power density. Accordingly, a good output of heating power occurs only in particle systems with narrow size and anisotropy distribution, as well as the mean diameter, being well adjusted in relation to the AC field frequency [326, 327]. In addition, the efficiency of heating is largely dependent on the ability of MNPs to specifically accumulate in the target tissues and to achieve effective cancer therapy with a minimum dosage.

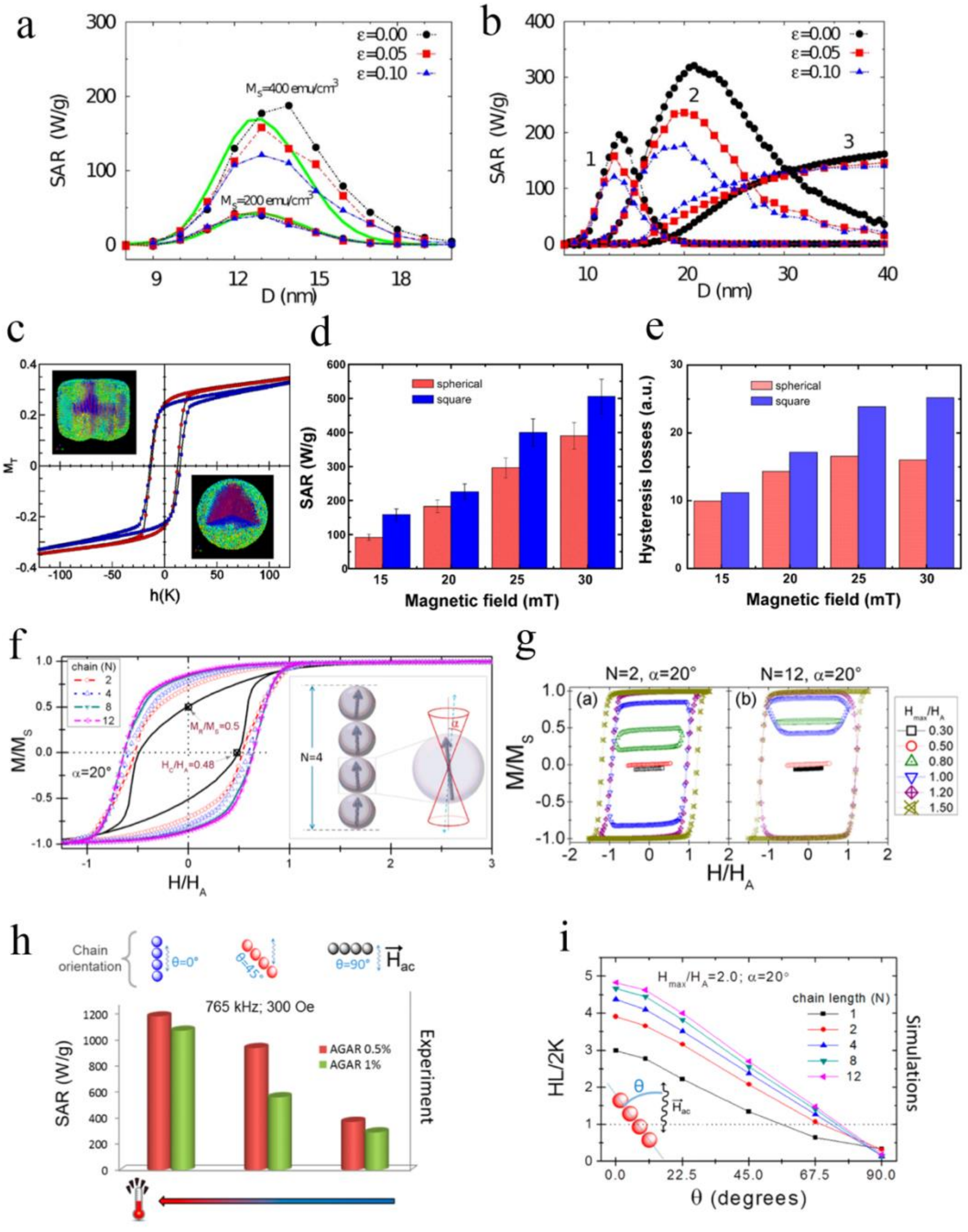

Figure 23. (a) SAR as a function of the mean particle size $D$ for variable volumetric packing fraction ε including non-interacting case (ε=0) and two values $M_s = 200\ emu/cm^3$ and $400\ emu/cm^3$. The solid line represents the solutions by means of RT. (b) SAR versus $D$ for different packing fractions and the values of $K = 3 \times 10^5\ erg/cm^3$ (curve set 1), $1.5 \times 10^5\ erg/cm^3$ (curve set 2) and $0.5 \times 10^5\ erg/cm^3$ (curves set 3). Calculations correspond to log-normal distributions of size and anisotropy constants both with standard deviations 0.1, spherically distributed anisotropy axes, and field amplitude $H = 300\ Oe$. (c) Hysteresis loops for a spherical

(red circles, diameter 20 nm) and a cubic particle (blue squares, side 20 nm) obtained from MC simulations of an atomistic spin model of maghemite at low temperature. In both, uniaxial anisotropy at the core and surface anisotropy have been considered. Spins have been colored according to their projection into the magnetic field direction (z-axis) from red (+1) to blue (-1). (d) & (e) SAR values for two nanoparticle solutions of similar concentration (0.5 mg/mL) and size volume but different shape indicating enhancement of SAR values for the 20 nm square nanoparticles. (d) and (e) are experimental and MC simulation results for the macrospin model with dipolar interactions at 300K. (f) Representative hysteresis M(H) normalized loops corresponding to different lengths (N) of a chain of magnetic nanoparticles with uniaxial easy anisotropy axes distributed within a cone of angle α = 20°. The black hysteresis curve corresponds to the case of randomly distributed non-interacting particles, also with the easy axes distributed at random. The inset illustrates the distribution of the axes within the cone of angle α, for the N = 4 chain. (g) The main panel shows the dependence of hysteresis losses $HL/2K$ vs $H_{max}$ for different chain lengths, for the same α = 20° case. Inset illustrates the optimal conditions for hyperthermia applications, as inferred by plotting $HL/2K \cdot H_{max}$ versus $H_{max}$. (h) Experimental hyperthermia power of the 0.5 % and 1 % agar cases measured at 765 kHz and 300 Oe maximum AC field, applied at different orientations with respect to the chains long axis (0°, 45°, and 90°; schematically illustrated in the top panel). (i) Simulated angle-dependence of the hysteresis losses for different chain lengths. The dotted line represents the hysteresis losses of a randomly distributed noninteracting system. ((a)&(b) from reference [328] under CC BY 4.0, (c)-(e) from reference [329] under CC BY 4.0, (f)-(i) reprinted with permission from reference [330], copyright 2014 American Chemical Society)

In recent years, many studies have been focused on optimizing the heating efficiency in terms of MNP intrinsic properties such as particle size [327, 329, 331, 332], anisotropy constant [329, 333-335], saturation magnetization [328], easy axis orientations [330, 336], extrinsic properties such as AC field frequency [327] and amplitude [329, 334], and the role of dipolar interactions [322, 329, 330, 337-339].

Ruta *et al.* [328] reported a kinetic Monte-Carlo (MC) method for the study of dipolar interaction, particle size, and saturation magnetization dependence of the SAR. The dependence of SAR on the particle diameter for two different values of $M_s$ is shown in Figure 23(a). For low $M_s = 200\ emu/cm^3$the dipolar interactions are weak and SAR does not depend on volumetric packing fraction ε. The MC calculation also extended toward the hysteretic regime in Figure 23(b), for small anisotropy constant $K$, increasing ε can lead to enhancement of SAR (see curve 3) while, on the other hand, for large $K$, the dipolar interactions are seen to decrease the SAR value (see curve 1). Martinez-Boubeta *et al.* [329] demonstrated that single-domain cubic iron oxide particles show superior magnetic heating efficiency compared to spherical particles of similar size, evidencing the beneficial role of surface anisotropy in the improved heating power. The observed area of the hysteresis loop of the cubic particle is bigger than that of spherical one as shown in Figure 23(c), which is further confirmed by experimental results in Figure 23(d) and MC simulation in Figure 23(e). On the other hand, the oriented attachment of MNPs into chain-like arrangements, biomimicking magnetotactic bacteria, have been recognized as an important pathway in the magnetic hyperthermia therapy roadmap. Serantes *et al.* [330] reported that the MNP chain has superior heating performance, increasing more than 5 times in comparison with the randomly distributed system when aligned with the magnetic field. In their study, the hysteresis cycles of an ensemble of non-interacting chains formed by N particles each were firstly investigated, a random angular distribution of anisotropy easy axis within the angle $\alpha$ that defines a cone around the longitudinal direction of the chain is assumed as shown in Figure 23(f). The role of chain length is illustrated in Figure 23(g), for the extreme cases of N = 2 and 12, respectively, the hysteresis area increases with increasing chain length, and it strongly depends on the filed amplitude $H_{max}$. Furthermore, they found that an increase of solution viscosity (as shown in Figure 23(h), agar > 0.5%) results in a decrease in the SAR. Heating power is also dependent on the relative orientation between AC field direction and chain, as shown in Figure 23(i).

## 7 MNPs for Bioseparation and Manipulation

### 7.1 Magnetic Separation

The separation and concentration of target analytes during sample preparation are primary steps in many biological studies such as disease diagnosis. Although magnetic diagnosis nowadays makes it possible to conduct bioassays on minimally processed biofluid samples, it remains a great challenge for the detection/quantification of analytes that are of very low-abundance due to the current instrumental detection limits and/or interferences from the complicated matrix. Based on this need, the sample preparation prior to an analytical process is introduced in order to improve the detection limit, this sample preparation step generally involves isolation of analytes from sample matrix, removal of interfering species and enrichment of analytes [46, 340-342]. Magnetic separation technique has several advantages in comparison with conventional separation methods such as ultrafiltration [343, 344], precipitation[345, 346], electrophoresis [347-350], etc. Furthermore, magnetic separation is usually gentle and nondestructive to biological analytes such as protein and peptides [351]. As shown in Figure 24(a), in a typical magnetic separation process, MNPs (or magnetic beads, MB) are employed to specifically conjugate to target analytes (cells, proteins, pathogenic substances, etc.), then the conjugated complexes can be easily and selectively removed by a subsequent magnetic separation process [352].

Magnetic separation technique has been applied in combination with the majority of other procedures used in biological analysis [353]. Chen *et al.* [212] reported a one-step sensing methodology that combines Magnetic Separation and Magnetic Relaxation Switching (MS-MRSw) assay for the detection of bacteria and viruses with high sensitivity and reproducibility. As shown in Figure 24(b), this method employs the different separation speeds of MB30 (30 nm magnetic bead) and MB250 (250 nm magnetic bead) in a small magnetic field (0.01 T), and the $T_2$ of the water molecules around the unreacted MB30 as the readout of the immunoassay. This MS-MRSw based assay reached a LOD of $10^2$ cfu/mL for the detection of *S. enterica*, compared to the conventional MRSw assay with a LOD of $10^4$ cfu/mL. Yang *et al.* [293] reported a sensing methodology that combines Magnetic Separation and Surface-enhanced Raman Scattering (MS-SERS). They have successfully applied this method for the detection of chloramphenicol (CAP) and reached a LOD of 1 pg/mL with a detection range of $1 - 10^4$ pg/mL.

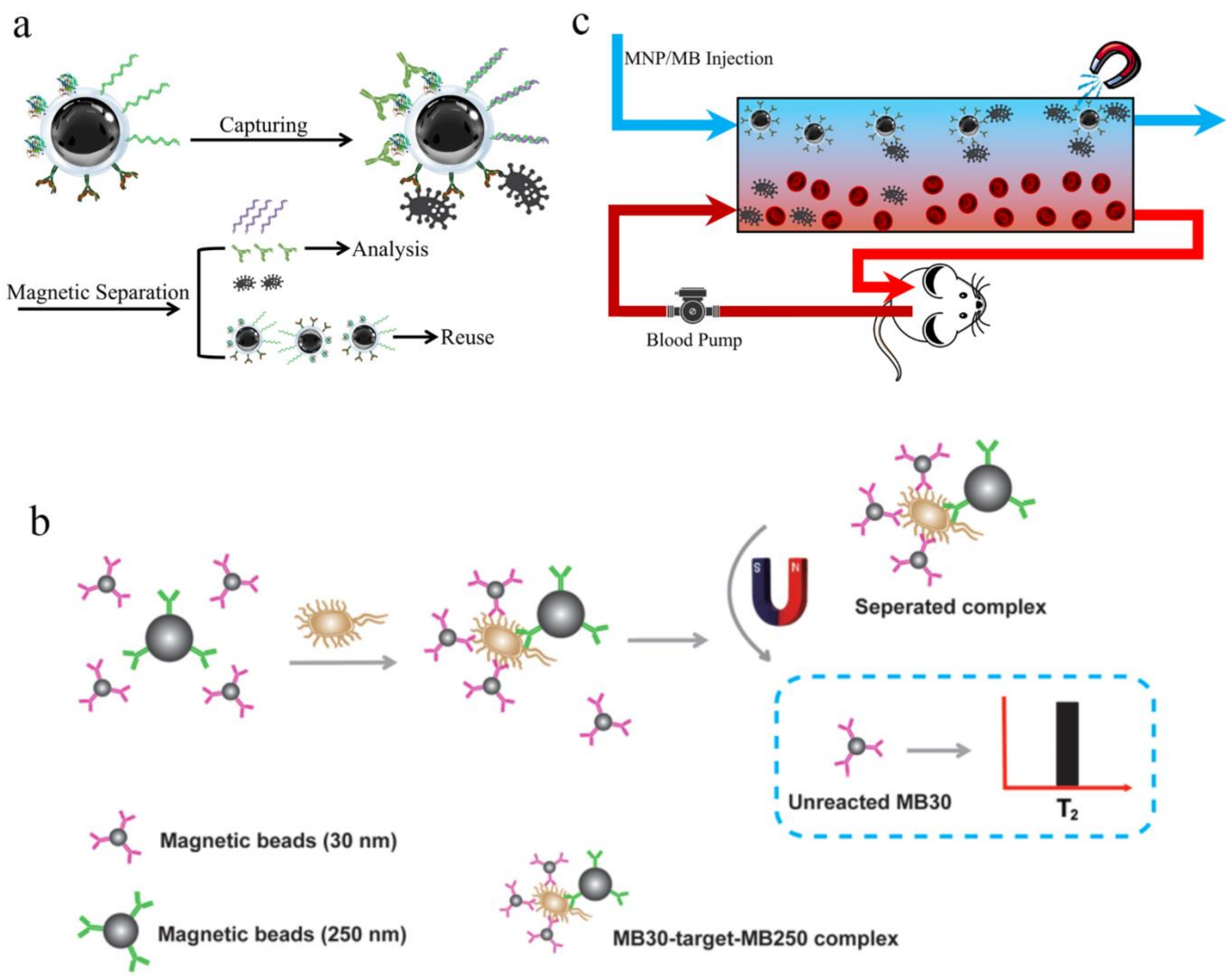

Figure 24. (a) The general procedure of magnetic separation in sample preparation for biological analysis. (b) Schematic illustration of the MS-MRSw sensor. MB250 and MB30 can selectively capture and enrich the target to form the sandwich "MB250-target-MB30" conjugate. After the magnetic separation, the $T_2$ signal of water molecules around the unreacted MB30 can be employed as the readout. (c) The principle of magnetic separation-based blood purification: elimination of pathogens. ((b) reprinted with permission from reference [212], copyright 2015 American Chemical Society)

## 7.2 Magnetic Manipulation

In addition to magnetic separation, this technique has also been applied to manipulate and/or sort cells, such as blood purification [47, 354-357], where the direct removal of disease-causing compounds is inherently attractive treatment modality for a range of pathological conditions such as bloodstream infections. In this magnetic separation-based blood purification process, MNPs/MBs conjugated with capture agents are injected into an extracorporeal blood circuit (see Figure 24(c)). Then the pathogen loaded MNPs/MNs are removed from the blood by magnetic separation. The effectiveness of blood purification is largely dependent on the target-ligand binding affinity.

# 8 Conclusions and Future Perspectives

The past three decades have been an exciting period in the synthesis of MNPs with interesting physical properties for their use in biological and biomedical applications, and many of these synthesis methods have been put in commercial production. To date, MNP is a burgeoning field as more improved techniques are being available for clinical therapy and diagnostics with increased sensitivity and efficiency. Furthermore, the surface functionalization of MNPs with polymers, biomolecules and ligands is crucial in order to impart biological recognition and interaction skills. Different kinds of MNPs have been developed to assist the bio-medical applications, such as MNPs with high magnetic moment, core@shell MNPs, core@shell@shell MNPs etc. The tunable properties due to the existence of the surface groups largely increases MNPs' compatibility with massive biomedical applications. As a matter of fact, it is also important that the toxicity of nanoparticles in biomedicine is also taken into consideration in future. It can be foreseen that in the near future the design of MNPs will revolutionize the medical healthcare field by their applications in the development of ultrasensitive and multiplexed diagnostic systems, *in vivo* imaging systems with high spatial and temporal resolutions, targeted and remotely controlled drug/gene delivery system for effective treatment of diseases, and gene therapy.

MR and μHall sensors represent another equally important example which is also a product of interdisciplinary associations. Besides, there are some very encouraging nanotechnological methods that include molecular-ruler technique, nanotransfer printing and nanoskiving that are yet to be used to fabricate magnetic structures. Although recent in origin, micromagnetic tools for simulation and modelling of magnetic materials have opened a new horizon of research that helps predicting properties of biosensors and their approximate results in their applications even before fabrication. This makes the total process all the more cost effective, error free and therefore saves time. It is rational to expect that the future progress in magnetism-based nanomedicine lies in the development of the latest technological advances in the fields of micro- and nano-fabrication. Apart from using these magnetic biosensors as portable lab-on-chip devices, the fact that they were integrated with IoT platforms, increased their commercial value to a great extent. This in turn had revolutionized the modern eHealth architecture.

## Author Information

To be filled.

## Acknowledgements

This study was financially supported by the Institute of Engineering in Medicine of the University of Minnesota, National Science Foundation MRSEC facility program, the Distinguished McKnight University Professorship, Centennial Chair Professorship, Robert F Hartmann Endowed Chair, and UROP program from the University of Minnesota.